\documentclass{article}
\usepackage[T1]{fontenc}
\usepackage[utf8]{inputenc}
\usepackage{color}
\usepackage{amsmath}
\usepackage{graphicx}
\usepackage{esint}
\usepackage{microtype}
\usepackage[unicode=true,
 bookmarks=false,
 breaklinks=false,pdfborder={0 0 1},backref=section,colorlinks=false]
 {hyperref}

\makeatletter
     \usepackage[nonatbib,preprint]{neurips_2022}

\usepackage{url}% simple URL typesetting
\usepackage{amsfonts}% blackboard math symbols
\usepackage{nicefrac}% compact symbols for 1/2, etc.
\usepackage{xcolor}% colors

\usepackage{mathtools}

\usepackage[nodisplayskipstretch]{setspace}
\newcommand\HSTeq{\stackrel{\mathclap{\normalfont\mbox{HST}}}{=}}

\title{A theory of learning with constrained weight-distribution}

\author{%
  Weishun Zhong \\
  Harvard and MIT\\
  \texttt{wszhong@mit.edu} \\
   \And
   Ben Sorscher \\
   Stanford University \\
   \texttt{bsorsch@stanford.edu} \\
   \AND
   Daniel D Lee \\
   Cornell Tech \\
   \texttt{ddl46@cornell.edu} \\
   \And
  Haim Sompolinsky \\
   Harvard and Hebrew University \\
   \texttt{haim@fiz.huji.ac.il} \\
}

\makeatother

\begin{document}
\maketitle 
\begin{abstract}
A central question in computational neuroscience is how structure
determines function in neural networks. Recent large-scale connectomic
studies have started to provide a wealth of structural information
such as the distribution of excitatory/inhibitory cell and synapse
types as well as the distribution of synaptic weights in the brains
of different species. The emerging high-quality large structural datasets
raise the question of what general functional principles can be gleaned
from them. Motivated by this question, we developed a statistical
mechanical theory of learning in neural networks that incorporates
structural information as constraints. We derived an analytical solution
for the memory capacity of the perceptron, a basic feedforward model
of supervised learning, with constraint on the distribution of its
weights. Interestingly, the theory predicts that the reduction in
capacity due to the constrained weight-distribution is related to
the Wasserstein distance between the cumulative distribution function
of the constrained weights and that of the standard normal distribution.
To test the theoretical predictions, we use optimal transport theory
and information geometry to develop an SGD-based algorithm to find
weights that simultaneously learn the input-output task and satisfy
the distribution constraint. We show that training in our algorithm
can be interpreted as geodesic flows in the Wasserstein space of probability
distributions.  We further developed a statistical mechanical theory
for teacher-student perceptron rule learning and ask for the best
way for the student to incorporate prior knowledge of the rule (i.e.,
the teacher). Our theory shows that it is beneficial for the learner
to adopt different prior weight distributions during learning, and
shows that distribution-constrained learning outperforms unconstrained
and sign-constrained learning. Our theory and algorithm provide novel
strategies for incorporating prior knowledge about weights into learning,
and reveal a powerful connection between structure and function in
neural networks. 
\end{abstract}

\section{Introduction}

Learning and memory are thought to take place at the microscopic level
by modifications of synaptic connections. Unlike learning in artificial
neural networks, synaptic plasticity in the brain operates under structural
biological constraints. Theoretical efforts to incorporate some of
these constraints have focused largely on the degree of connectivity
\cite{buzsaki2014log,koulakov2009correlated} and the constraints
on the sign of the synapses (Excitatory vs. Inhibitory) \cite{amit1989perceptron,brunel2004optimal},
but few include additional features of synaptic weight distributions
observed in the brain \cite{barbour2007can}. More generally, recent
large-scale connectomic studies \cite{kunst2019cellular,scheffer2020connectome,shapson2021connectomic}
are beginning to provide a wealth of structural information of neuronal
circuits at an unprecedented scope and level of precision, which presents
a remarkable opportunity for a more refined theoretical study of learning
and memory that takes into account these hitherto unavailable structural
information.

\begin{figure}
\centering{}\includegraphics[scale=0.15]{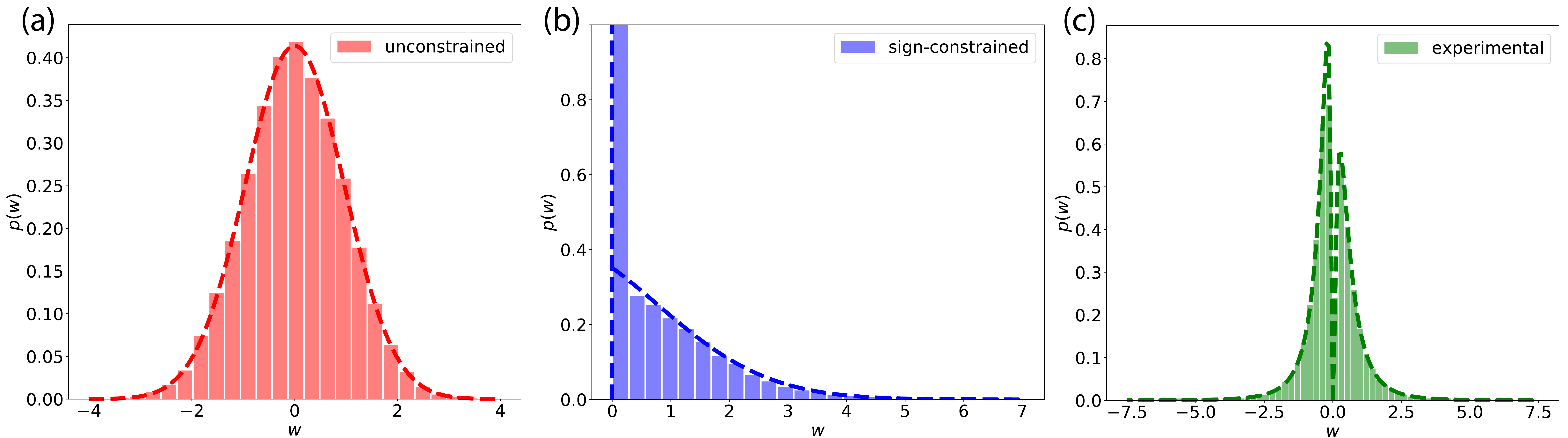}\caption{\label{fig:motivation}Theoretical and empirical synaptic weight distributions.
(a)-(b) predicted distribution following perceptron learning at capacity.
(a) Normal distribution when learning is unconstrained. (b) A delta-function
plus a half-normal distribution when learning is sign-constrained.
(c) Experimentally measured synaptic weight distribution (mouse primary
auditory cortex \cite{levy2012spatial}).}
\end{figure}

Perceptron \cite{rosenblatt1958perceptron} is arguably the simplest
model of computation by single neuron and is the fundamental building
block for many modern neural networks. Despite the drastic oversimplification,
studying the computational properties of (binary and analog) perceptron
has been used extensively in computational neuroscience since its
dawn, particularly in the cerebellum (as a model of sensory-motor
association) but also in cerebral cortex (for generic associative
memory functions) \cite{marr1969theory,albus1971theory,brunel2004optimal,chapeton2012efficient,brunel2016cortical,bouvier2018cerebellar}.
Forming associations is considered an ‘atomic’ building block for
generic cortical functions, and perceptron memory capacity sets a
tight bound on the memory capacity in recurrently connected neuronal
circuits with application to cortex and hippocampus\textcolor{red}{{}
}\cite{gardner1987maximum,rolls1998neural,rubin2017balanced}. Statistical
mechanical analysis predicts that near capacity, an unconstrained
perceptron classifying random input-output associations has normally
distributed weights \cite{gardner1988optimal,gardner1988space,4038449},
see Fig.\ref{fig:motivation}(a). In contrast, physiological experiments
suggest that biological synapses do not change their excitatory/inhibitory
identity during learning (but see recent \cite{kim2022co}). In order
to take perceptron a step closer to biological realism, prior work
has imposed sign constraints during learning \cite{amit1989perceptron,brunel2004optimal}.
In this case, the predicted weight distribution is a delta-function
centered at zero plus a half-normal distribution, see Fig.\ref{fig:motivation}(b).
However, a wide range of connectomic studies ranging from cortical
circuits in animals \cite{levy2012spatial,holmgren2003pyramidal,molnar2008complex,yang2013development,shapson2021connectomic,loewenstein2011multiplicative,avermann2012microcircuits},
to human cerebral cortex \cite{molnar2008complex,shapson2021connectomic}
have shown evidence of lognormally distributed synaptic connections.
As an example, Fig.\ref{fig:motivation}(c) shows the weight connection
distribution in mouse primary auditory cortex (data adapted from \cite{levy2012spatial}).
 Possible reasons for the ubiquitous lognormal distributions range
from biological structural/developmental constraints to computational
benefits \cite{teramae2014computational}. Various potential mechanisms
for lognormal distributions has been proposed, from multiplicative
gradient updates in feedforward networks\cite{kivinen1997exponentiated,loewenstein2011multiplicative},
to mixture of additive and multiplicative plasticity rules in spiking
networks\cite{gilson2011stability}, but the majority of these proposals
lead not just to lognormal distributions but also to sparsification
in the weights. Instead of adding yet another explanation to the
computational origin of lognormal distribution, here we take the observed
weight distribution as a prior on the network structure, and ask for
its computational consequences. The goal of the paper is to present
for the first time a quantitative and qualitative theory of neural
network learning performance under non-Gaussian and general weight
distributions (not limited to lognormal distributions). 

In this paper, we combine two powerful tools: statistical mechanics
and optimal transport theory, and present a theory of perceptron learning
that incorporates the knowledge of both distribution and sign information
as constraints, and gives accurate predictions for capacity and generalization
error. Interestingly, the theory predicts that the reduction in capacity
due to the constrained weight-distribution is related to the Wasserstein
distance between the cumulative distribution function of the constrained
weights and that of the standard normal distribution. Along with the
theoretical framework, we also present a learning algorithm derived
from information geometry that is capable of efficiently finding viable
perceptron weights that satisfy desired distribution and sign constraints.
This paper is organized as follows: in Section \ref{sec:capacity_one_population}
we derive the perceptron capacity for classifying random input-output
associations using statistical mechanics, and illustrate our theory
with a simple example. In Section \ref{sec:disco_algorithm}, we derive
our learning algorithm using optimal transport theory, and show that
distribution of weights found by the learning algorithm coincide with
geodesic distributions on a Wasserstein statistical manifold, and
therefore training can be interpreted as a geodesic flow. In Section
\ref{sec:compare_experiment} we analyze a parameterized family of
biologically realistic weight distributions, and use our theory to
predict the shape of the distribution with optimal parameters. We
map out the experimental parameter landscape for the estimated distribution
of synaptic weights in mammalian cortex and show that our theory's
prediction for optimal distribution is close to the experimentally
measured value. In Section \ref{sec:Generalization} we further develop
a statistical mechanical theory for teacher-student perceptron rule
learning and ask for the best way for the student to incorporate prior
knowledge about the weight distribution of the rule (i.e., the teacher).
Our theory shows that it is beneficial for the learner to adopt different
prior weight distributions during learning. 

\section{Capacity}

\subsection{Learning under weight distribution constraints}

\label{sec:capacity_one_population}

We begin by considering a canonical learning problem: classifying
random input-output associations by a perceptron. In biological memory
systems, the heavily correlated sensory data is undergoing heavy preprocessing
including massive decorrelations, and previous work on brain related
perceptron modeling \cite{gardner1987maximum,brunel2004optimal,rubin2017balanced}
assumes similarly unstructured data. The data consists of pairs$\left\{ \boldsymbol{\xi}^{\mu},\zeta^{\mu}\right\} _{\mu=1}^{P}$,
where $\boldsymbol{\xi}^{\mu}$ is an $N$-dimentional random vector
drawn i.i.d. from a standard normal distribution, $p(\xi_{i}^{\mu})=\mathcal{N}(0,1)$,
and $\zeta^{\mu}$ are random binary class labels with $p(\zeta^{\mu})=\frac{1}{2}\delta(\zeta^{\mu}+1)+\frac{1}{2}\delta(\zeta^{\mu}-1)$.
The goal is to find a hyperplane through the origin, described by
a perceptron weight vector $\boldsymbol{w}\in\mathbb{R}^{N}$, normalized
to $||\boldsymbol{w}||^{2}=N.$

We call $\boldsymbol{w}$ a separating hyperplane when it correctly
classifies all the examples with margin $\kappa>0$:

\begin{equation}
\zeta^{\mu}\frac{\boldsymbol{w}\cdot\boldsymbol{\xi}^{\mu}}{||\boldsymbol{w}||}\geq\kappa.\label{eq:margin}
\end{equation}

We are interested in solutions $\boldsymbol{w}$ to Eqn.\ref{eq:margin}
that obey a prescribed distribution constraint, $w_{i}\sim q(w)$,
where $q$ is an arbitrary probability density function. We further
demand that $\langle w^{2}\rangle_{q(w)}=1$ to fix the overall scale
of the distribution (since the task is invariant to the overall scale
of $w$). Thus, the goal of learning is to find weights that satisfy
\ref{eq:margin} with the additional constraint that the empirical
density function $\hat{q}(w)=\frac{1}{N}\sum_{i}^{N}\delta(w-w_{i})$,
formed by the learned weights is similar to $q(w)$, and more precisely
that it converges to $q(w)$ as $N\rightarrow\infty$ (see Section
\ref{subsec:capacity_gardner} below). 

Extension of this setup that includes an arbitrary number of populations
each satisfying its own prescribed distribution constraints is discussed
in Section \ref{sec:compare_experiment} and in Appendix \ref{app:capacity_M_pop}.
Note that the sign constraint is a special case of this scenario with
two synaptic populations: one excitatory and one inhibitory. We further
discuss the generalization of this setup to include biased inputs
and sparse labels in Appendix \ref{app:capacity_biased_sparse}.

\subsection{Statistical mechanical theory of capacity}

\label{subsec:capacity_gardner}

We are interested in the thermodynamic limit where $P,N\to\infty$
, but the load $\alpha=\frac{P}{N}$ stays $\mathcal{O}(1)$. This
limit is amenable to mean-field analysis using statistical mechanics.

Following Gardner's seminal work \cite{gardner1988optimal,gardner1988space},
we consider the fraction $V$ of viable weights that satisfies both
Eqn.\ref{eq:margin} and the distribution constraint $\hat{q}=q$$,$
to all possible weights:

\begin{equation}
V=\frac{\int d\boldsymbol{w}\left[\prod_{\mu=1}^{P}\Theta\left(\zeta^{\mu}\frac{\boldsymbol{w}\cdot\boldsymbol{\xi}^{\mu}}{||\boldsymbol{w}||}-\kappa\right)\right]\delta(||\boldsymbol{w}||^{2}-N)\delta\bigg(\int dk\left(\hat{q}(k)-q(k)\right)\bigg)}{\int d\boldsymbol{w}\delta(||\boldsymbol{w}||^{2}-N)}.\label{eq:volume}
\end{equation}
In Eqn.\ref{eq:volume}, we impose the distribution constraint $\hat{q}=q$
by demanding that in the thermodynamic limit, all Fourier modes of
$q$ and $\hat{q}$ are the same , i.e., that $q(k)=\int dwe^{ikw}q(w)$
= $\hat{q}(k)=\frac{1}{N}\sum_{i}^{N}e^{ikw_{i}},$where in the last
equality we have used the definition of empirical distribution. We
perform a quenched average over random patterns $\boldsymbol{\xi}^{\mu}$
and labels $\zeta^{\mu}$. This amounts to calculating $\left\langle \log V\right\rangle $,
which can be done using the replica trick \cite{gardner1988optimal,gardner1988space}.

We focus on solutions with maximum margin $\kappa$ at a given load
$\alpha$, or equivalently, the maximum load capacity $\alpha_{c}(\kappa)$
of separable patterns given margin $\kappa$. We proceed by assuming
replica symmetry in our mean field analysis, which in general might
not hold because the constraint $\hat{q}=q$ is non-convex. For all
the results presented in the main text, replica symmetry solution
is supported by numerical simulations. In Appendix \ref{app:Replica-symmetry-breaking}
we explore the validity of replica symmetric solutions in the case
of strongly bimodal distributions and show that they fail only very
close to the binary (Ising) limit.

Detailed calculations of the mean-field theory are presented in Appendix
\ref{app:capacity_dist_const}. Our mean-field theory predicts that
the reduction in capacity due to the distribution constraint is proportional
to the Jacobian of the transformation from $w\sim q(w)$ to a normally
distributed variable $x(w)\sim\mathcal{N}(0,1)$,

\begin{equation}
\alpha_{c}(\kappa)=\alpha_{0}(\kappa)\left\langle \frac{dw}{dx}\right\rangle _{x}^{2},\label{eq:single_pop}
\end{equation}

where $\alpha_{0}(\kappa)=\left[\int_{-\kappa}^{\infty}Dt(\kappa+t)^{2}\right]^{-1}$
is the capacity of an unconstrained perceptron, from Gardner theory
\cite{gardner1988optimal,gardner1988space}, and $\kappa=0$ reduces
to the classical result of $\alpha_{0}(0)=2.$ The Jacobian factor,
$\langle dw/dx\rangle_{x}$, can be written in terms of the constrained
distribution's cumulative distribution function (CDF), $Q(w)$, and
the standard normal CDF $P(x)=\frac{1}{2}\left[1+\text{Erf}(\frac{x}{\sqrt{2}})\right]$,
namely,

\begin{equation}
\left\langle \frac{dw}{dx}\right\rangle _{x}=\int_{0}^{1}duQ^{-1}(u)P^{-1}(u).\label{eq:jacobian}
\end{equation}

Note that since the second moments are fixed to unity, $0\leq\left\langle \frac{dw}{dx}\right\rangle _{x}\leq1$
and it equals $1$ iff $p=q.$ 

\begin{figure}
\centering{}\includegraphics[scale=0.15]{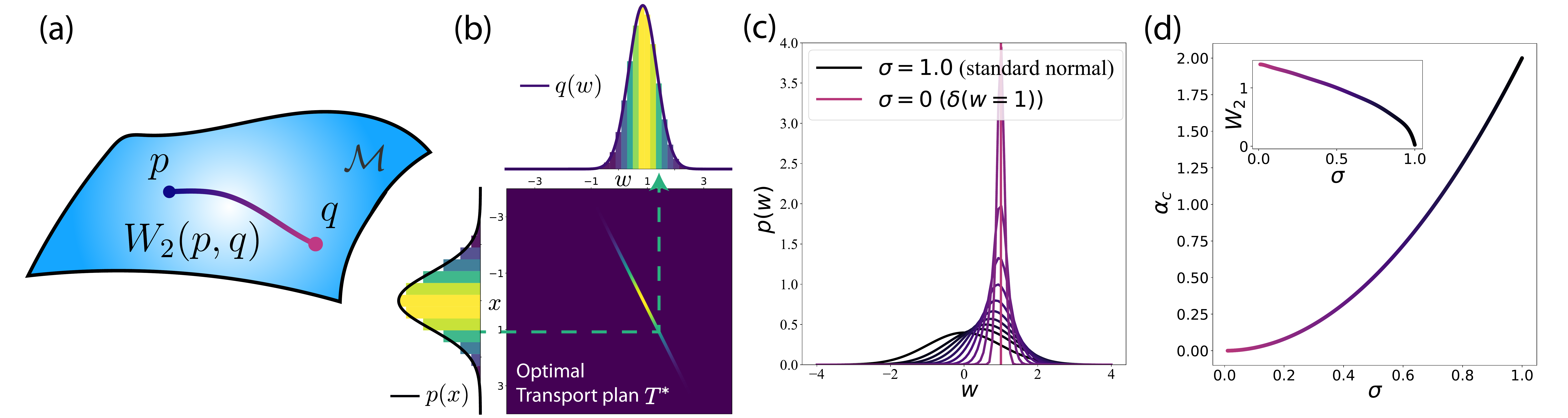}\caption{\label{fig:theory_illustration}An illustration of optimal transport
from a standard normal distribution $\mathcal{N}(0,1)$ to normal
distributions with nonzero mean $\mathcal{N}(\sqrt{1-\sigma^{2}},\sigma^{2})$.
(a) A schematic of the space $(\mathcal{M},W_{2})$ of probability
distributions. (b) An example optimal transport plan from standard
normal, $p(x)$, to a normal with $\sigma=0.5$, $q(w)$. The optimal
transport plan $T^{*}$ is plotted in between the distributions. $T^{*}$
moves $p(x)$ units of probability mass $x$ to location $w$, as
indicated by the dashed line, and the colors are chosen to reflect
the amount of probability mass to be transported. (c) $\mathcal{N}(\sqrt{1-\sigma^{2}},\sigma^{2})$
interpolates between standard normal ($\sigma=0$) to a $\delta$-function
at 1 ($\sigma=1$). (d) Capacity $\alpha_{c}(\kappa=0)$ as a function
of $\sigma$. Inset shows the $W_{2}$ distance as a function of $\sigma$.}
\end{figure}

\subsection{Geometrical interpretation of capacity}

\label{subsec:Geometrical-interpretation}

The jacobian factor Eqn.\ref{eq:jacobian} can be rewritten as

\begin{equation}
\left\langle \frac{dw}{dx}\right\rangle _{x}=1-\frac{1}{2}W_{2}(Q,P)^{2},\label{eq:W2_dist}
\end{equation}
where $W_{k}$ ($k=2$ in above) is the Wasserstein-$k$ distance,
given by

\begin{equation}
W_{k}(Q,P)=\left[\int_{0}^{1}du\left(Q^{-1}(u)-P^{-1}(u)\right)^{k}\right]^{1/k}.\label{eq:Wk_dist}
\end{equation}

{[}In the following, we will make frequent use of both the probability
density function (PDF), and the cumulative distribution function (CDF).
We distinguish them by using upper case letters for CDFs, and lower
case letters for PDFs.{]} 

The Wasserstein distance measures the dissimilarity between two probability
distributions, and is the geodesic distance between points on the
manifold of probability distributions \cite{lott2006some,figalli2011optimal,chen2020optimal}.
Therefore, we can interpret Eqn.\ref{eq:single_pop} as predicting
that the reduction in memory capacity tracks the geodesic distance
we need travel from the standard normal distribution $P$ to the target
distribution $Q$ (Fig.\ref{fig:theory_illustration}(a)).

We demonstrate Eqn.\ref{eq:single_pop} and Eqn.\ref{eq:W2_dist}
with an instructive example. Let's consider a parameterized family
of normal distributions, with the second moment fixed to 1: $q(w)=\mathcal{N}(\sqrt{1-\sigma^{2}},\sigma^{2})$,
see Fig.\ref{fig:theory_illustration}(c). At $\sigma=1$, $q(w)$
is the standard normal distribution and we recover the unconstrained
Gardner capacity $\alpha_{0}(\kappa=0)=2$. As $\sigma\to0,$ $q(w)$
becomes a $\delta$-function at $1$ and $\alpha_{c}(\kappa)\to0$
(Fig.\ref{fig:theory_illustration}(c)). 

As evident in this simple example, perceptron capacity is strongly
affected by its weight distribution. Our theory enables prediction
of the shape of the distribution with optimal parameters within a
parameterized family of distributions. We apply our theory to a family
of biologically plausible distributions and compare our prediction
with experimentally measured distributions in Section \ref{sec:compare_experiment}.

\section{Optimal transport and the DisCo-SGD learning algorithm}

\label{sec:disco_algorithm}

Eqn.\ref{eq:single_pop} predicts the storage capacity for a perceptron
with a given weight distribution, but it does not specify a learning
algorithm for finding a solution to this non-convex learning problem.
Here we present a learning algorithm for perceptron learning with
a given weight distribution constraint. This algorithm will also serve
to test our theoretical predictions. For this purpose, we use optimal
transport theory to develop an SGD-based algorithm that is able to
find max-margin solutions that obey the prescribed distribution constraint.
Furthermore, we show that training can be interpreted as traveling
along the geodesic connecting the current empirical distribution and
the target distribution. 

Stochastic gradient descent (SGD) on a cross-entropy loss has been
shown to asymptotically converge to max-margin solutions on separable
data \cite{soudry2018implicit,nacson2019stochastic}. Given data $\left\{ \boldsymbol{\xi}^{\mu},\zeta^{\mu}\right\} _{\mu=1}^{P}$,
we use logistic regression to predict class labels from our perceptron
weights, $\hat{\zeta}^{\mu}=\sigma(\boldsymbol{w}^{t}\cdot\boldsymbol{\xi}^{\mu}),$where
$\sigma(z)=\left(1+e^{-z}\right)^{-1}$ and $\boldsymbol{w}^{t}$
is the weight at the $t$-th update. This defines an SGD update rule
:

\begin{equation}
w_{i}^{t+\delta t}\leftarrow w_{i}^{t}-\delta t\sum_{\mu}\xi_{i}^{\mu}(\hat{\zeta}^{\mu}-\zeta^{\mu}),\label{eq:SGD}
\end{equation}

where the $\mu$-summation goes from $1$ to $P$ for full-batch GD
and goes from $1$ to mini-batch size $B$ for mini-batches SGD (see
Appendix \ref{app:DiscoSGD} for more details). The theory of optimal
transport provides a principled way of transporting each individual
weight $w_{i}^{t}$ to a new value so that overall the new set of
weights satisfies the prescribed target distribution. In $1$-D, the
optimal transport plan $T^{*}$ has a closed-form solution in terms
of the current CDF $P$ and target CDF $Q$ \cite{thorpe2019introduction,ambrosio2013user}:
$T^{*}=Q^{-1}\circ P$, where $\circ$ denotes functional composition.
We demonstrate the optimal transport map in Fig.\ref{fig:theory_illustration}(b)
for the instructive example discussed in Section \ref{subsec:Geometrical-interpretation}. 

\begin{table}
\centering{}\includegraphics[scale=0.22]{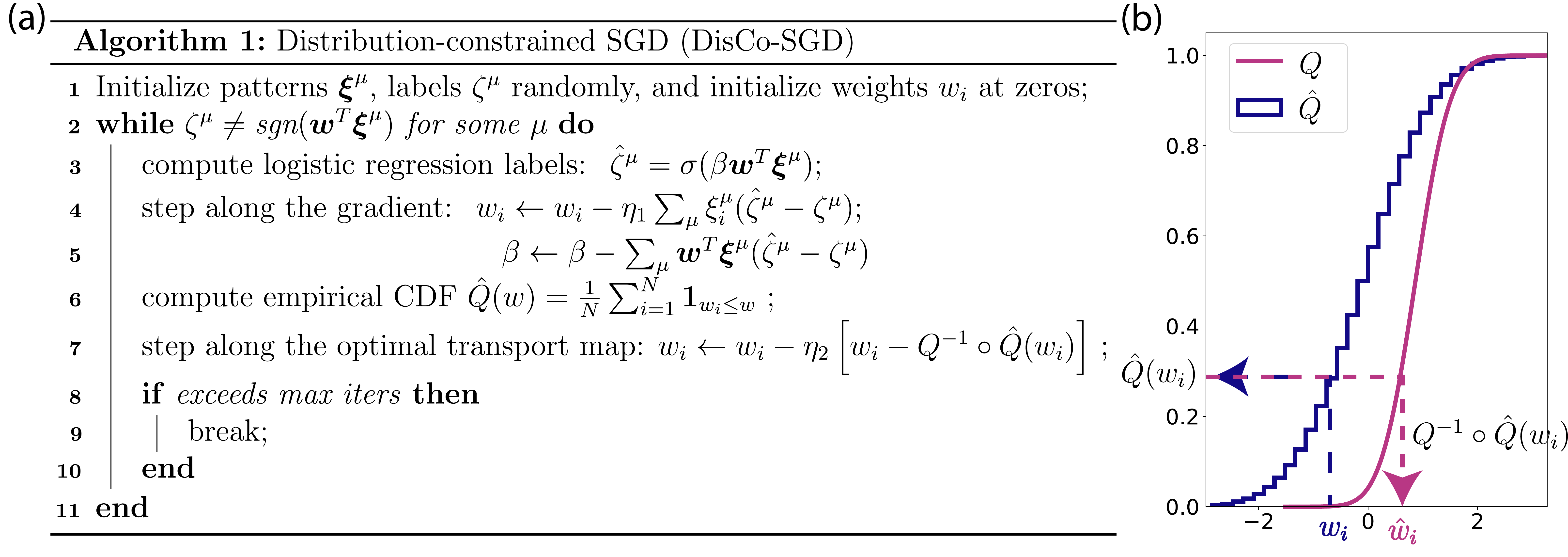}\caption{\label{tab:discoSGD}Disco-SGD algorithm. (a) We perform alternating
steps of gradient descent along the cross-entropy loss (Eqn.\ref{eq:SGD}),
followed by steps along the optimal transport direction (Eqn.\ref{eq:soft_constraint}).
(b) An illustration of Eqn.\ref{eq:w_hat}. For a given $w_{i}$,
we first compute its empirical CDF value $\hat{Q}(w_{i}),$then use
the inverse target CDF to transport $w_{i}$ to its new value, $\hat{w_{i}}=Q^{-1}\left(\hat{Q}(w_{i})\right)$.}
\end{table}

In order to apply $T^{*}$ to transport our weights $\left\{ w_{i}\right\} $
(omitting superscript $t$), we form the empirical CDF $\hat{Q}(w)=\frac{1}{N}\sum_{i=1}^{N}\mathbf{1}_{w_{i}\leq w}$,
which counts how many weights $w_{i}$ are observed below value $w$.
Then the new set of weights $\left\{ \hat{w}_{i}\right\} $ satisfying
target CDF $Q$ can be written as 

\begin{equation}
\hat{w}_{i}=Q^{-1}\circ\hat{Q}(w_{i}).\label{eq:w_hat}
\end{equation}

We illustrate Eqn.\ref{eq:w_hat} in action in Table \ref{tab:discoSGD}(b).

\begin{figure}
\centering{}\includegraphics[scale=0.3]{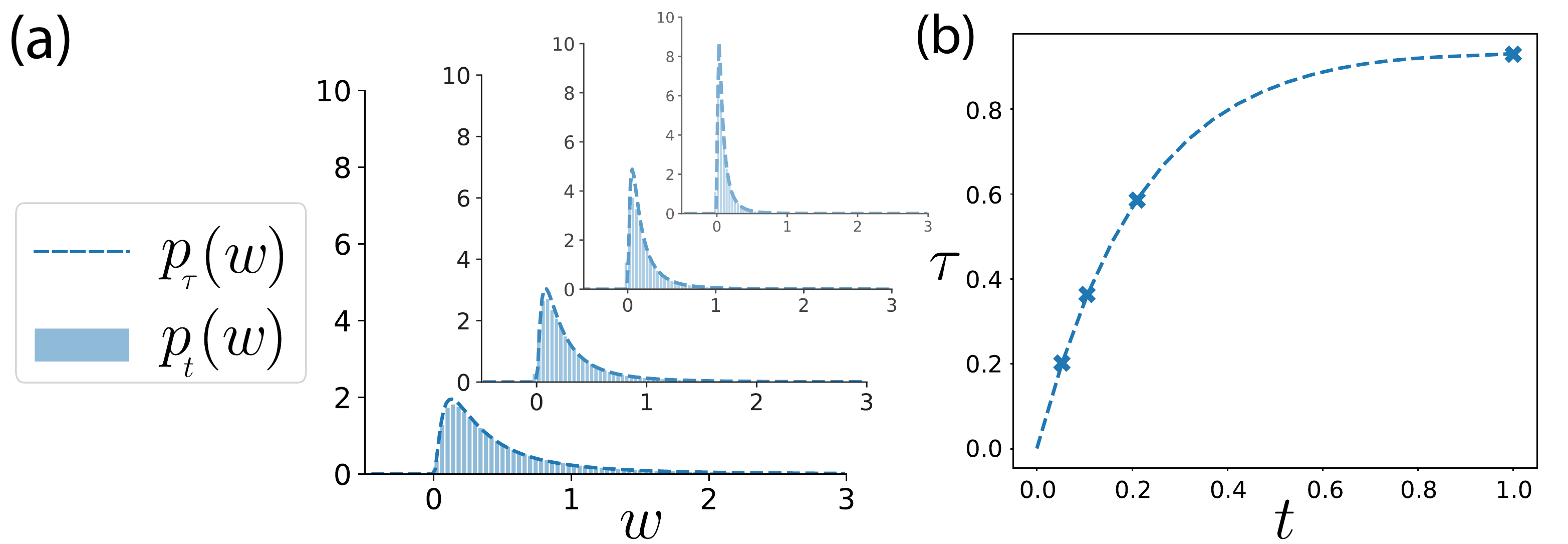}\caption{\label{fig:geodesic}Intermediate distributions during learning are
on the geodesic. (a) The solid histograms are the intermediate distribution
$p_{t}$ at different training time $t$ from the DisCo-SGD algorithm,
the dashed lines are geodesic distributions $p_{\tau}$ with the same
$W_{2}$ distance to the target distribution $Q$. From right to left
the training time advances, and the distributions transform further
away from the $\delta$-function initialization, and approach the
target distribution (a lognormal, in this example). (b) The geodesic
time $\tau$ as a function of the training time $t$. Location of
the crosses correspond to the distributions shown in (a).}
\end{figure}

However, performing such a one-step projection strongly interferes
with the cross-entropy objective, and numerically often results in
solutions that do not perfectly classify the data. Therefore, it would
be beneficial to have an incremental update rule based on Eqn.\ref{eq:w_hat}:

\begin{equation}
w_{i}^{\tau+\delta\tau}\leftarrow w_{i}^{\tau}+\delta\tau\left(\hat{w}_{i}-w_{i}^{\tau}\right),\label{eq:soft_constraint}
\end{equation}

where we have used a different update time $\tau$ to differentiate
with the cross-entropy update time $t$. 

We present our complete algorithm in Table \ref{tab:discoSGD}(a),
which we named `Distribution-constrained SGD' (DisCo-SGD) algorithm.
In the DisCo-SGD algorithm, we perform alternating updates on Eqn.\ref{eq:SGD}
and Eqn.\ref{eq:soft_constraint}, and identify $\delta t$ and $\delta\tau$
as learning rates $\eta_{1}$ and $\eta_{2}$. Note that in logistic
regression, the norm of the weight vector $||\boldsymbol{w}||$ is
known to increase with training and the max-margin solution is only
recovered at $||\boldsymbol{w}||\to\infty$. In contrast, imposing
a distribution constraint fixes the norm. Therefore, to allow a variable
norm, in Table \ref{tab:discoSGD} we include a trainable parameter
$\beta$ in our algorithm to serve as the norm of the weight vector.
This algorithm allows us to reliably discover linearly separable solutions
obeying the prescribed weight distribution $Q$.

Interestingly, Eqn.\ref{eq:soft_constraint} takes a similar form
to geodesic flows in Wasserstein space. Given samples $\left\{ w_{i}\right\} $
drawn from the initial distribution $P$ and $\left\{ \hat{w}_{i}\right\} $
drawn from the final distribution $Q$, samples $\left\{ w_{i}^{\tau}\right\} $
from intermediate distributions $P_{\tau}$ along the geodesic can
be calculated as $w_{(i)}^{\tau}=(1-\tau)w_{(i)}+\tau\hat{w}_{(i)}$,
where subscript $(i)$ denotes ascending order (see more in Appendix
\ref{app:Optimal-transport-theory}). For intermediate perceptron
weights $\boldsymbol{w}^{t}$ found by our algorithm, we can compute
its empirical distribution $p_{t}$ and compare with theoretical distribution
$p_{\tau}$ along the geodesic with the same $W_{2}$ distance to
the target distribution (see Appendix \ref{app:Optimal-transport-theory}
for how to calculate $p_{\tau}$). In Fig.\ref{fig:geodesic}(a),
we show that indeed the empirical distributions $p_{t}$ agree with
the geodesic distributions $p_{\tau}$ at geodesic time $\tau(t)$
(Fig.\ref{fig:geodesic}(a)). The relation between the geodesic time
$\tau$ and the SGD update time $t$ is shown in Fig.\ref{fig:geodesic}(b).
The interplay between the cross-entropy objective and the distribution
constraint is manifested in the rate at which the distribution moves
along the geodesic between the initial distribution and the target
one. 

\begin{figure}
\centering{}\includegraphics[scale=0.25]{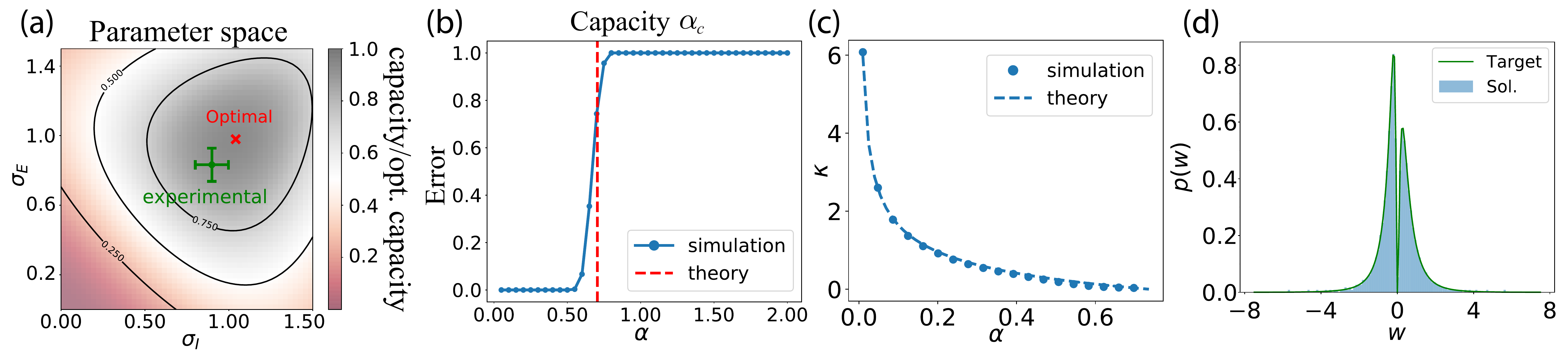}\caption{\label{fig:Experimental-landscape}Biologically-realistic distribution
and parameter landscape. (a) Capacity (normalized by the optimal value
in the landscape) as a function of the lognormal parameters $\sigma_{E}$
and $\sigma_{I}$. Experimental value is shown in green with error
bars, and optimal capacity is shown in red. (b)-(d) (theory from Eqn.\ref{eq:capacity_M_pop}
and simulations from DisCo-SGD): (b) Determination of capacity; (c)
Max-margin $\kappa$ at different load $\alpha$, which is the same
as $\alpha_{c}(\kappa)$; (d) Example weight distribution obtained
in simulation.}
\end{figure}

\section{Biologically-realistic distribution (E/I balanced lognormals) and
experimental landscape}

\label{sec:compare_experiment}

In order to apply our theory to the more biologically-realistic cases,
we generalize our theory from a single prescribed distribution to
an arbitrary number of input subpopulations each obeys its own distribution.
We consider a perceptron that consists of $M$ synaptic populations
$\boldsymbol{w}^{m}$ indexed by $m$, each constrained to satisfy
its own weight distribution $w_{i}^{m}\sim q_{m}(w^{m})$. We denote
the overall weight vector as $\boldsymbol{w}\equiv\{\boldsymbol{w}^{m}\}_{m=1}^{M}\in\mathbb{R}^{N\times1}$,
where the total number of weights is $N=\sum_{m=1}^{M}N_{m}$. In
this case, the capacity Eqn.\ref{eq:single_pop} is generalized to
(See Appendix \ref{app:capacity_M_pop} for detailed derivation):

\begin{equation}
\alpha_{c}(\kappa)=\alpha_{0}(\kappa)\left[\sum_{m}^{M}g_{m}\left\langle \frac{dw^{m}}{dx}\right\rangle _{x}\right]^{2},\label{eq:capacity_M_pop}
\end{equation}

where $g_{m}=N_{m}/N$ is the fraction of weights in this population.
Eqn. \ref{eq:capacity_M_pop} allows us to investigate the parameter
space of capacity with biologically-realistic distributions and compare
with the experimentally measured values. In particular, we are interested
the case with two synaptic populations that models the excitatory/inhibitory
synpatic weights of a biological neuron, hence, $m=E,I$. We model
the excitatory/inhibitory synaptic weights as drawn from two separate
lognormal distributions $(g_{I}=1-g_{E}$): $w_{i}^{E}\sim\text{\ensuremath{\frac{1}{\sqrt{2\pi}\sigma_{E}w^{E}}\exp\left\{ -\frac{(\ln w^{E}-\mu_{E})^{2}}{2\sigma_{E}^{2}}\right\} }}$
and $w_{i}^{I}\sim\text{\ensuremath{\frac{1}{\sqrt{2\pi}\sigma_{I}w^{I}}\exp\left\{ -\frac{(\ln w^{I}-\mu_{I})^{2}}{2\sigma_{I}^{2}}\right\} }}.$

We also demand that the mean synaptic weights satisfy the E/I balance
condition \cite{van1996chaos,van1998chaotic,tsodyks1995rapid,van2005irregular,rubin2017balanced,mongillo2018inhibitory,chapeton2012efficient}
$g_{E}\left\langle w^{E}\right\rangle =g_{I}\left\langle w^{I}\right\rangle $
as is often observed in cortex connectomic experiments \cite{anderson2000orientation,wehr2003balanced,okun2008instantaneous,poo2009odor,atallah2009instantaneous}.
With the E/I balance condition and fixed second moment, the capacity
is a function of the lognormal parameters $\sigma_{E}$ and $\sigma_{I}$.
In Fig.\ref{fig:Experimental-landscape}(a) we map out the 2d parameter
space of $\sigma_{E}$ and $\sigma_{I}$ using Eqn.\ref{eq:capacity_M_pop},
and find that the optimal choice of parameters which yields the maximum
capacity solution is close to the experimentally measured values in
a recent connectomic studies in mouse primary auditory cortex \cite{levy2012spatial}. 

In order to test our theory's validity on this estimated distribution
of synaptic weights, we perform DisCo-SGD simulation with model parameters
$\sigma_{E}$ and $\sigma_{I}$ fixed to their experimentally measured
values. Both the capacity (Fig.\ref{fig:Experimental-landscape}(b)),
max-margin $\kappa$ at different load (Fig.\ref{fig:Experimental-landscape}(c)),
and the empirical weights found by the algorithm (Fig.\ref{fig:Experimental-landscape}(d))
are in good agreement with our theoretical prediction.

\section{Generalization performance}

\label{sec:Generalization}

\subsection{Distribution-constrained learning as circuit inference}

A central question in computational neuroscience is how underlying
neural circuits determine its computation. Recently, thanks to new
parallelized functional recording technologies, simultaneous recordings
of the activity of hundreds of neurons in response to an ensemble
of inputs are possible \cite{ahrens2013whole,berenyi2014large}. An
interesting challenge is to infer the structural connectivity from
the measured input-output activity patterns. It is interesting to
ask how are these stimuli-response relations related to the underlying
structure of the circuit \cite{real2017neural,liu2017inference}.
In the following, we try to adress this circuit reconstruction task
in a simple setup where a student perceptron tries to learn from a
teacher perceptron \cite{seung1992statistical,engel2001statistical}.
In this setup, the teacher is considered to be the underlying ground-truth
neural circuit. The student is attempting to infer the connection
weights of this ground-truth circuit by observing a series of input-output
relations generated by the teacher. After learning is completed, one
can assess the faithfulness of the inference by comparing the teacher
and student. The teacher-student setup is also a well-known ‘toy model’
for studying generalization performance\textcolor{red}{{} }\cite{loureiro2021learning,lee2021continual,matiisen2019teacher}.
In this case since the learning data are generated by the teacher,
the overlap between teacher and student determines the generalization
performance of the learning. Here we ask to what extent prior knowledge
of the teacher weight distribution helps in learning the rule and
how this knowledge can be incorporated in learning. A similar motivation
may arise in other contexts, in which there is a prior knowledge about
the weight distribution of an unknown target linear classifier. 

Let's consider the teacher perceptron, $\boldsymbol{w}_{t}\in\mathbb{R}^{N}$,
drawn from some ground-truth distribution $p_{t}$. Given random inputs
$\boldsymbol{\xi}^{\mu}$ with $p(\xi_{i}^{\mu})=\mathcal{N}(0,1)$,
we generate labels by $\zeta^{\mu}=\text{sgn}(\boldsymbol{w}_{t}\cdot\boldsymbol{\xi}^{\mu}/||\boldsymbol{w}_{t}||+\eta^{\mu})$,
where $\eta^{\mu}$ is input noise and $\eta^{\mu}\sim\mathcal{N}(0,\sigma^{2})$.
We task the student perceptron $\boldsymbol{w}_{s}$ to find the max-margin
linear classifier for data $\{\boldsymbol{\xi}^{\mu},\zeta^{\mu}\}_{\mu=1}^{p}$:
$\max\kappa:\zeta^{\mu}\boldsymbol{w}_{s}\cdot\boldsymbol{\xi}^{\mu}\geq\kappa||\boldsymbol{w}_{s}||$.
Let's define the teacher-student overlap as

\begin{equation}
R=\frac{\boldsymbol{w}_{s}\cdot\boldsymbol{w}_{t}}{\left\Vert \boldsymbol{w}_{s}\right\Vert \left\Vert \boldsymbol{w}_{t}\right\Vert },\label{eq:overlap}
\end{equation}

which is a measure the faithfulness of the circuit inference. The
student's generalization error is then related to the overlap by $\varepsilon_{g}=1/\pi\arccos\left(R/\sqrt{1+\sigma^{2}}\right)$
\cite{seung1992statistical,engel2001statistical}.

\begin{figure}
\centering{}\includegraphics[scale=0.2]{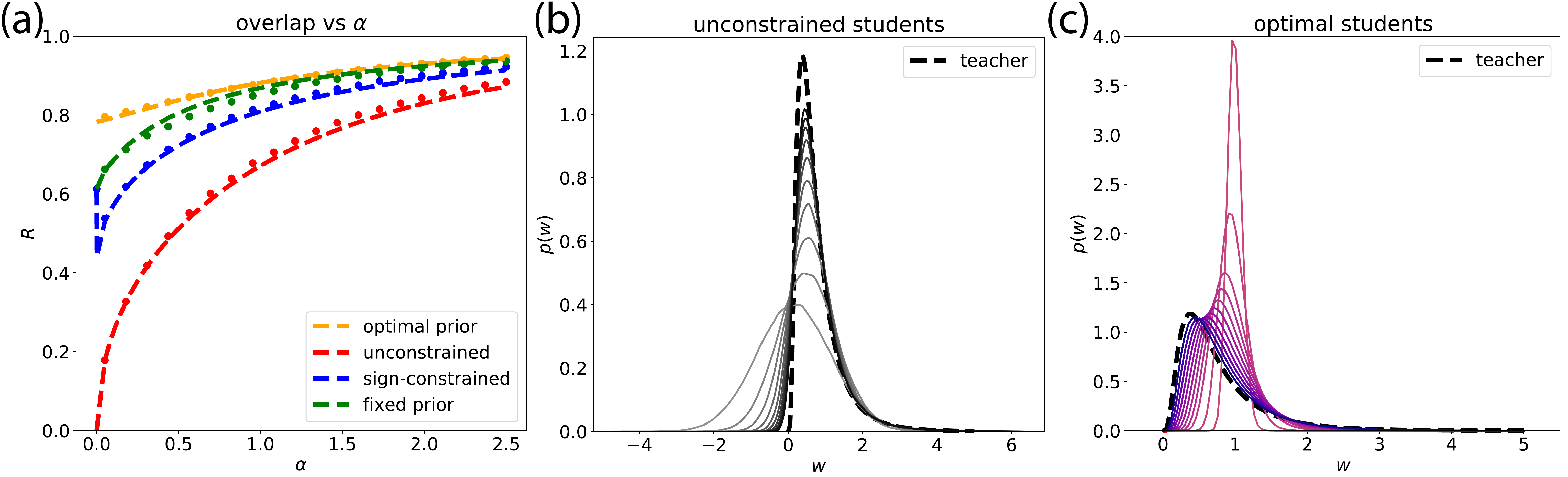}\caption{\label{fig:compare_learnings}Compare different learning paradigms.
(a) Teacher-student overlap $R$ , or equivalently the generalization
error $\varepsilon_{g}=1/\pi\arccos R$, as a function of load $\alpha$
in different learning paradigms. Dashed lines are from theory, and
dots are from simulation. Note that there is an initial drop of the
overlap in sign-constrained learning due to sparsification of weights.
(b)-(c) The darker color curves correspond to larger $\alpha$, and
dashed line is teacher distribution (same in both cases). (b) Distribution
of an unconstrained student evolves from normal distribution toward
the teacher distribution. (c) Optimal student prior evolves from a
$\delta$-function toward the teacher distribution.}
\end{figure}

As a baseline, let's first consider a totally uninformed student (without
any structural knowledge of the teacher), learning from a teacher
with a given (in particular non-Gaussian) weight distribution. In
this case, we can determine the overlap $R$ (Eqn.\ref{eq:overlap})
as a function of load $\alpha$ by solving the replica symmetric mean
field self-consistency equations as in \cite{seung1992statistical,engel2001statistical}.
An example of such learning for a lognormal teacher distribution is
shown in Fig.\ref{fig:compare_learnings}(a) (`unconstrained') for
the noiseless case ($\sigma=0$). Note that in the presence of noise
in the labels $(\sigma\neq0)$, $\alpha$ is bounded by $\alpha_{c}(\sigma)$
, since max-margin learning of separable data is assumed. The case
with nonzero $\sigma$ is presented in Appendix \ref{app:generalization_noise}.
In this unconstrained case, the student's weight distribution evolves
from a Gaussian for low $\alpha$ to one which increasingly resembles
the teacher distribution for large $\alpha$ (Fig.\ref{fig:compare_learnings}(b)). 

Next, we consider a student with information about the signs of the
individual teacher weights. We can apply this knowledge as a constraint
and demand that the signs of individual student weights agree with
that of the teacher's. The additional sign-constraints require a modification
of replica calculation in \cite{seung1992statistical,engel2001statistical},
which we present in Appendix \ref{app:generalization_sign_const}.
Surprisingly, we find both analytically and numerically that if the
teacher weights are not too sparse, the max-margin solution generalizes
poorly: after a single step of learning (with random input vectors),
the overlap, $R$, drops substantially from its initial value (see
`sign-constrained' in Fig.\ref{fig:compare_learnings}(a)). The source
of the problem is that, due to the sign constraint, max-margin training
with few examples yields a significant mismatch between the student
and teacher weight distributions. After only a few steps of learning,
half of the student's weights are set to zero, and the student's distribution,
$p(w_{s})=1/2\delta(0)+1/\sqrt{2\pi}\exp\{-w_{s}^{2}/4\}$, deviates
significantly from the teacher's distribution (see more in Appendix
\ref{app:generalization_sparsification}). The discrepancy between
the teacher and student weight distributions therefore suggest that
we should incorporate distribution-constraint into learning.

\subsection{Distribution-constrained learning outperforms unconstrained and sign-constrained
learning }

Let's consider the case that the student weight are constrained to
some \textit{prior} distribution $q_{s}(w_{s})$, while the teacher
obeys a distribution $p_{t}(w_{t}),$for an arbitrary pair $q_{s},p_{t}$.
We can write down the Gardner volume $V_{g}$ for generalization as
in the capacity case (Eqn.\ref{eq:volume}):

\begin{equation}
V_{g}=\frac{\int d\boldsymbol{w}_{s}\left[\prod_{\mu=1}^{P}\Theta\left(\text{sgn}\left(\frac{\boldsymbol{w}_{t}\cdot\boldsymbol{\xi}^{\mu}}{||\boldsymbol{w}_{t}||}+\eta^{\mu}\right)\frac{\boldsymbol{w}_{s}\cdot\boldsymbol{\xi}^{\mu}}{||\boldsymbol{w}_{s}||}-\kappa\right)\right]\delta(||\boldsymbol{w}_{s}||^{2}-N)\delta\bigg(\int dk\left(\hat{q}(k)-q(k)\right)\bigg)}{\int d\boldsymbol{w}_{s}\delta(||\boldsymbol{w}_{s}||^{2}-N)}.\label{eq:volume-generalization}
\end{equation}

To obtain ensemble average of system over different realizations of
the training set, we perform the quenched average of $\log V_{g}$
over the patterns $\boldsymbol{\xi}^{\mu}$ and teacher $\boldsymbol{w}_{t}$,
and consider the thermodynamic limit of $N,P\to\infty$ and $\alpha=\frac{P}{N}$
stays $\mathcal{O}(1)$. We use the replica trick similar to \cite{seung1992statistical,engel2001statistical}.
Overlap $R$ (Eqn.\ref{eq:overlap}) can be determined as a function
of load $\alpha$ by solving the replica symmetric mean field self-consistency
equations in Appendix \ref{app:generalization_dist_const}. In this
distribution-constrained setting, we can perform numerical simulations
with DisCo-SGD algorithm (Table \ref{tab:discoSGD}) to find such
weights and compare with the predictions of our theory. 

Now we ask if the student has a \textit{prior} on the teacher's weight
distribution $p_{t}$, whether incorporating this knowledge in training
will improve generalization performance. One might be tempted to conclude
that the optimal prior distribution the student should adopt is always
that of the teacher's, i.e., $q_{s}=p_{t}$. We call this learning
paradigm `fixed prior', and show that its generalization performance
is better than that of the unconstrained and sign-constrained case
(Fig.\ref{fig:compare_learnings}(a)). However, instead of using a
fixed prior for the student, we can in fact choose the \textit{optimal
prior} distribution $p_{s}^{*}$ at different load $\alpha$. This
presents a new learning paradigm we called `optimal prior'. In Fig.\ref{fig:compare_learnings}(a),
we show that choosing optimal priors at different $\alpha$ achieves
the overall best generalization performance compared with all other
learning paradigms. For a given parameterized family of distributions,
our theory provides a way to analytically obtain the optimal prior
$p_{s}^{*}$ as a function of $\alpha$ (Fig.\ref{fig:compare_learnings}(c)).
Note that unlike the unconstrained case (Fig.\ref{fig:compare_learnings}(b)),
the optimal prior starts from a $\delta$-function at $1$ at zero
$\alpha$, and asymptotically approaches the teacher distribution
$p_{t}$ as $\alpha\to\infty$.

\section{Summary and Discussion}

We have developed a statistical mechanical framework that incorporates
structural constraints (sign and weight distribution) into perceptron
learning. The synaptic weights in our perceptron learning satisfy
two key biological constraints: (1) individual synaptic signs are
not affected by the learning task (2) overall synaptic weights obey
a prescribed distribution. These constraints may arise also in neuromorphic
devices \cite{han2021alternative,truong2014new}. Under the replica-symmetry
assumption, we derived a novel form of distribution-constrained perceptron
storage capacity, which admits a simple geometric interpretation of
the reduction in capacity in terms of the Wasserstein distance between
the standard normal distribution and the imposed distribution. To
numerically test our analytic theory, we used tools from optimal transport
and information geometry to develop an SGD-based algorithm, DisCo-SGD,
in order to reliably find weights that satisfy such prescribed constraints
and correctly classify the data, and showed that training with the
algorithm can be interpreted as geodesic flows in the Wasserstein
space of distributions. It would be interesting to compare our theory
and algorithm to \cite{arjovsky2017wasserstein,sanjabi2018convergence}
where the Wasserstein distance is used as an objective for training
generative models. We applied our theory to the biologically realistic
case of of excitatory/inhibitory lognormal distributions that are
observed in the cortex, and found experimentally-measured parameters
close to the optimal parameter values predicted by our theory. We
further studied input-output rule learning where the target rule is
defined in terms of a weighted sum of the inputs, and asked to what
extent prior knowledge of the target distribution may improve generalization
performance. Using the teacher-student perceptron learning setup,
we showed analytically and numerically that distribution constrained
learning substantially enhances the generalization performance. In
the context of circuit inference, distribution constrained learning
provides a novel and reliable way to recover the underlying circuit
structure from observed input-output neural activities. In summary,
our work provides new strategies of incorporating knowledge about
weight distribution in neural learning and reveals a powerful connection
between structure and function in neural networks. Ongoing extensions
of the present work include weight distribution constraints in recurrent
and deep architectures as well as testing against additional connectomic
databases. 

\section*{Acknowledgments}

This paper is dedicated to the memory of Mrs. Lily Safra, a great
supporter of brain research. The authors would like to thank Madhu
Advani, Haozhe Shan, and Julia Steinberg for very helpful discussions.
W.Z. acknowledges support from the Swartz Program in Theoretical Neuroscience
at Harvard and the NIH grant NINDS (1U19NS104653). B.S. acknowledges
support from the Stanford Graduate Fellowship. D.D.L. acknowledges
support from the NIH grant NINDS (1U19NS104653). H.S. acknowledges
support from the Swartz Program in Theoretical Neuroscience at Harvard,
the NIH grant NINDS (1U19NS104653), and the Gatsby Charitable Foundation. 

\section*{\medskip{}
 }

%%%%%%%%%%%%%%%%%%%%%%%%%%%%%%%%%%%%%%%%%%%%%%%%%%%%%%%%%%%%
\newpage

\appendix
%dummy comment inserted by tex2lyx to ensure that this paragraph is not empty%dummy comment inserted by tex2lyx to ensure that this paragraph is not empty%dummy comment inserted by tex2lyx to ensure that this paragraph is not empty

\section{Appendix}

\subsubsection*{Preliminaries}

Throughout the appendix, we make frequent use of Gaussian integrals.
We introduce short-hand notations $\int Dt\equiv\int\frac{dt}{\sqrt{2\pi}}e^{-t^{2}/2}$
and $H(x)\equiv\int_{x}^{\infty}Dt$. Also, when we do not specify
the integration range it is understood that we are integrating from
$-\infty$ to $\infty$. 

\subsection{Capacity supplemental materials}

\label{app:capacity}

\subsubsection{Replica calculation of distribution-constrained capacity}

\label{app:capacity_dist_const}

In this section, we present the replica calculation of the distribution-constrained
storage capacity of a perceptron. 

As described in main text Eqn.2, we need to perform a quenched average
$\left\langle \cdot\right\rangle $ over the patterns $\boldsymbol{\xi}^{\mu}$
and labels $\zeta^{\mu}$ for $\log V$, which can be carried out
using the replica trick, $\left\langle \log V\right\rangle =\lim_{n\to0}(\left\langle V^{n}\right\rangle -1)/n$.
Following \cite{gardner1988optimal,gardner1988space}, we consider
first integer $n$, and at the end perform analytic continuation of
$n\to0$. The replicated Gardner volume is:

\begin{equation}
V=\frac{\prod_{\alpha=1}^{n}\int d\boldsymbol{w}^{\alpha}\left[\prod_{\mu=1}^{P}\Theta\left(\zeta^{\mu}\frac{\boldsymbol{w}^{\alpha}\cdot\boldsymbol{\xi}^{\mu}}{||\boldsymbol{w}^{\alpha}||}-\kappa\right)\right]\delta(||\boldsymbol{w}^{\alpha}||^{2}-N)\delta\left(\int dk\bigg(\hat{q}(k)-q(k)\bigg)\right)}{\prod_{\alpha=1}^{n}\int d\boldsymbol{w}^{\alpha}\delta(||\boldsymbol{w}^{\alpha}||^{2}-N)}\label{appeq:volume_cap_dist}
\end{equation}

Let's rewrite the Heaviside step function using Fourier representation
of the $\delta$-function $\delta(x)=\int_{-\infty}^{\infty}\frac{dk}{2\pi}e^{ikx}$
as (defining $z_{\alpha}^{\mu}=\zeta^{\mu}\frac{\boldsymbol{w}^{\alpha}\cdot\boldsymbol{\xi}^{\mu}}{||\boldsymbol{w}^{\alpha}||}$)

\begin{equation}
\Theta\left(z_{\alpha}^{\mu}-\kappa\right)=\int_{\kappa}^{\infty}d\rho_{\alpha}^{\mu}\delta(\rho_{\alpha}^{\mu}-z_{\alpha}^{\mu})=\int_{\kappa}^{\infty}d\rho_{\alpha}^{\mu}\int\frac{dx_{\alpha}^{\mu}}{2\pi}e^{ix_{\alpha}^{\mu}(\rho_{\alpha}^{\mu}-z_{\alpha}^{\mu})}.
\end{equation}

Note that now all the $\boldsymbol{\xi}^{\mu},\zeta^{\mu}$ dependence
is in $e^{-ix_{\alpha}^{\mu}z_{\alpha}^{\mu}}$. We perform the average
with respect to $\xi_{i}^{\mu}\sim p(\xi_{i}^{\mu})=\mathcal{N}(0,1)$
and $p(\zeta^{\mu})=\frac{1}{2}\delta(\zeta^{\mu}+1)+\frac{1}{2}\delta(\zeta^{\mu}-1)$
(also note that $||\boldsymbol{w}^{\alpha}||=\sqrt{N}$):

\begin{equation}
\begin{split}\left\langle \prod_{\mu\alpha}e^{-ix_{\alpha}^{\mu}z_{\alpha}^{\mu}}\right\rangle _{\xi\eta} & =\prod_{\mu j}\left\langle \exp\left\{ -\frac{i}{\sqrt{N}}\zeta^{\mu}\xi_{j}^{\mu}\sum_{\alpha}x_{\alpha}^{\mu}w_{j}^{\alpha}\right\} \right\rangle _{\xi\zeta}\\
 & =\prod_{\mu i}\left\langle \exp\left\{ -\frac{(\zeta^{\mu})^{2}}{2N}\sum_{\alpha\beta}x_{\alpha}^{\mu}x_{\beta}^{\mu}w_{i}^{\alpha}w_{i}^{\beta}\right\} \right\rangle _{\zeta}\\
 & =\prod_{\mu}\exp\left\{ -\frac{1}{2N}\sum_{\alpha\beta}x_{\alpha}^{\mu}x_{\beta}^{\mu}\sum_{i}w_{i}^{\alpha}w_{i}^{\beta}\right\} .
\end{split}
\label{appeq:cap_dist_energy_x_z}
\end{equation}

Introducing the replica overlap parameter $q_{\alpha\beta}=\frac{1}{N}\sum_{i}w_{i}^{\alpha}w_{i}^{\beta}$,
and notice that the $\mu$ index gives $P$ identical copies of the
same integral. We can suppress the $\mu$ indices and write

\begin{equation}
\begin{split}\left\langle \prod_{\mu\alpha}\Theta\left(z_{\alpha}^{\mu}-\kappa\right)\right\rangle _{\xi\zeta} & =\left[\int_{\kappa}^{\infty}\left(\prod_{\alpha}\frac{d\rho_{\alpha}dx_{\alpha}}{2\pi}\right)e^{K}\right]^{P}\end{split}
,
\end{equation}

where

\begin{equation}
K=i\sum_{\alpha}x_{\alpha}\rho_{\alpha}-\frac{1}{2}\sum_{\alpha\beta}q_{\alpha\beta}x_{\alpha}x_{\beta}\label{appeq:cap_dist_K_replica}
\end{equation}

captures all the data dependence in the quenched free energy landscape,
and therefore it is called the `energetic' part of the free energy.
In contrast, the $\delta$-functions in Eqn.\ref{appeq:volume_cap_dist}
are called `entropic' part because they regulate what kind of weights
are considered in the version space (space of viable weights).

\subsubsection*{The entropic part}

\begin{equation}
\begin{split}\delta(Nq_{\alpha\beta}-\sum_{i}w_{i}^{\alpha}w_{i}^{\beta}) & =\int\frac{d\hat{q}_{\alpha\beta}}{2\pi}\exp\left\{ iN\hat{q}_{\alpha\beta}q_{\alpha\beta}-i\hat{q}_{\alpha\beta}\sum_{i}w_{i}^{\alpha}w_{i}^{\beta}\right\} \end{split}
.
\end{equation}

Note that the normalization constraint $\delta(||\boldsymbol{w}^{\alpha}||^{2}-N)$
is automatically satisfied by requiring $q_{\alpha\alpha}=1$. Using
replica-symmetric ansatz: $\hat{q}_{\alpha\beta}=-\frac{i}{2}(\Delta\hat{q}\delta_{\alpha\beta}+\hat{q}_{1})$,
and $q_{\alpha\beta}=(1-q)\delta_{\alpha\beta}+q$, we have

\begin{equation}
iN\sum_{\alpha\beta}\hat{q}_{\alpha\beta}q_{\alpha\beta}=\frac{nN}{2}\left[\Delta\hat{q}+\hat{q}_{1}(1-q)\right]+\mathcal{O}(n^{2}).
\end{equation}

and

\begin{equation}
\begin{split}-i\sum_{\alpha\beta}\hat{q}_{\alpha\beta}\sum_{i}w_{i}^{\alpha}w_{i}^{\beta} & =-\frac{1}{2}(\Delta\hat{q}+\hat{q}_{1})\sum_{\alpha}\sum_{i}(w_{i}^{\alpha})^{2}-\frac{1}{2}\hat{q}_{1}\sum_{(\alpha\beta)}\sum_{i}w_{i}^{\alpha}w_{i}^{\beta}\\
 & =-\frac{1}{2}\Delta\hat{q}\sum_{\alpha}\sum_{i}(w_{i}^{\alpha})^{2}-\frac{1}{2}\hat{q}_{1}\sum_{i}\left(\sum_{\alpha}w_{i}^{\alpha}\right)^{2}\\
 & \HSTeq-\frac{1}{2}\Delta\hat{q}\sum_{\alpha}\sum_{i}(w_{i}^{\alpha})^{2}+\sqrt{-\hat{q}_{1}}\sum_{i}t_{i}\left(\sum_{\alpha}w_{i}^{\alpha}\right),
\end{split}
\end{equation}

where in the last step HST denotes Hubbard-Stratonovich transformation
$\int\frac{dt}{\sqrt{2\pi}}e^{-t^{2}/2}e^{bt}=e^{b^{2}/2}$ that we
use to linearize the quadratic term at the cost of introducing an
auxiliary Gaussian variable $t$ to be averaged over later.

Recall that $\hat{q}(k)=\int e^{ikw}\hat{p}(w)=\frac{1}{N}\sum_{i}^{N}e^{ikw_{i}^{\alpha}}$,
the distribution constraint becomes 

\begin{equation}
\begin{split}\delta\bigg(\int dk\left(\hat{q}(k)-q(k)\right)\bigg) & =\delta\left(\int dk\left(\frac{1}{N}\sum_{i}^{N}e^{ikw_{i}^{\alpha}}-q(k)\right)\right)\\
 & =\int\frac{d\hat{\lambda}_{\alpha}(k)}{2\pi}\exp\left\{ \int dki\hat{\lambda}_{\alpha}(k)\left(\sum_{i}e^{ikw_{i}^{\alpha}}-Nq(k)\right)\right\} .
\end{split}
\label{appeq:cap_dist_dist_const_delta}
\end{equation}

Note that $\sum_{i}\int dki\hat{\lambda}_{\alpha}(k)e^{ikw_{i}^{\alpha}}=2\pi i\sum_{i}\lambda_{\alpha}(-w_{i}^{\alpha})$
by inverse Fourier transform. Next,

\begin{equation}
\begin{split}-iN\int dk\hat{\lambda}_{\alpha}(k)q(k)= & -iN\int dk\left(\int dwe^{ikw}\lambda_{\alpha}(w)\right)\left(\int dw'e^{ikw'}q(w')\right)\\
 & =-2\pi iN\int dwdw'\lambda_{\alpha}(w)q(w')\delta(w+w')\\
 & =-2\pi iN\int dwq(w)\lambda_{\alpha}(-w).
\end{split}
\end{equation}

Now we can write down the full free energy. We ignore overall constant
coefficients such as $2\pi$'s and $i$'s in the integration measure,
which become irrelevant upon taking the saddle-point approximation.
We also leave out the denominator of $V$, as it does not depend on
data and is an overall constant. Note that under the replica-symmetric
ansatz the replica index $\alpha$ gives $n$ identical copies of
the same integral and thus the replica index $\alpha$ can be suppressed
(same for synaptic index $i$):

\begin{equation}
\left\langle V^{n}\right\rangle =\int dqd\hat{\lambda}(k)d\Delta\hat{q}d\hat{q}_{1}e^{nN(G_{0}+G_{1})},\label{appeq:cap_dist_free_energy}
\end{equation}

where (please note that $q$ is replica overlap, and $q(w)$ is the
imposed target distribution)

\begin{equation}
\begin{split}G_{0} & =\frac{1}{2}\Delta\hat{q}+\frac{1}{2}\hat{q}_{1}(1-q)-2\pi i\int dwq(w)\lambda(-w)+\left\langle \log Z(t)\right\rangle _{t},\\
Z(t) & =\int dw\exp\left\{ 2\pi i\lambda(-w)-\frac{1}{2}\Delta\hat{q}w^{2}+\sqrt{-\hat{q}_{1}}tw\right\} .
\end{split}
\end{equation}

Note that integrals in Eqn.\ref{appeq:cap_dist_free_energy} can be
evaluated using saddle-point approximation in the thermodynamic limit
$N\to\infty$.

Redefining $2\pi i\lambda(-w)-\frac{1}{2}\Delta\hat{q}w^{2}\to-\lambda(w)$
and $-\hat{q}_{1}\to\hat{q}_{1}$, we have 

\begin{equation}
\begin{split}G_{0} & =\frac{1}{2}\Delta\hat{q}-\frac{1}{2}\hat{q}_{1}(1-q)+\int dwq(w)\lambda(w)-\frac{1}{2}\Delta\hat{q}\int dwq(w)w^{2}+\left\langle \log Z(t)\right\rangle _{t},\\
Z(t) & =\int dw\exp\left\{ -\lambda(w)+\sqrt{\hat{q}_{1}}tw\right\} .
\end{split}
\label{appeq:G0_cap_dist_qneq1}
\end{equation}

We seek the saddle-point solution for $G_{0}$ with respect to the
order parameters $\Delta\hat{q}$, $\lambda(w)$, and $\hat{q}_{1}$:

\begin{equation}
\begin{split}0=\frac{\partial G_{0}}{\partial\Delta\hat{q}} & \Rightarrow1=\int dwq(w)w^{2}=\left\langle w^{2}\right\rangle _{q(w)},\end{split}
\label{appeq:second_moment}
\end{equation}
\begin{equation}
\begin{split}0=\frac{\partial G_{0}}{\partial\lambda(w)} & \Rightarrow q(w)=\left\langle \frac{1}{Z(t)}\exp\left\{ -\lambda(w)+\sqrt{\hat{q}_{1}}tw\right\} \right\rangle \end{split}
.\label{appeq:saddle_pt_1_qneq1}
\end{equation}

We observe that the saddle-point equation Eqn.\ref{appeq:second_moment}
fixes the second moment of the imposed distribution $q(w)$ to 1 and
therefore can be thought of as a second moment constraint. $G_{0}$
now simplifies to

\begin{equation}
G_{0}=-\frac{1}{2}\hat{q}_{1}(1-q)+\int dwq(w)\lambda(w)+\left\langle \log Z(t)\right\rangle _{t}.
\end{equation}

The remaining $\hat{q}_{1}$ saddle-point equation is a bit more complicated,

\begin{equation}
\begin{split}0=\frac{\partial G_{0}}{\partial\hat{q}_{1}} & =-\frac{1}{2}(1-q)+\frac{t}{2\sqrt{\hat{q}_{1}}}\left\langle \frac{1}{Z(t)}\int dww\exp\left\{ -\lambda(w)+\sqrt{\hat{q}_{1}}tw\right\} \right\rangle _{t}\end{split}
\end{equation}

Integration by parts for the second term in rhs:
\begin{equation}
\begin{split}1-q= & \frac{1}{\sqrt{\hat{q}_{1}}}\int Dt\frac{1}{Z}\sqrt{\hat{q}_{1}}\int dww^{2}\exp\left\{ -\lambda(w)+\sqrt{\hat{q}_{1}}tw\right\} \\
- & \frac{1}{\sqrt{\hat{q}_{1}}}\int Dt\frac{1}{Z^{2}}\sqrt{\hat{q}_{1}}\left(\int dww\exp\left\{ -\lambda(w)+\sqrt{\hat{q}_{1}}tw\right\} \right)^{2}\\
= & \left\langle \left\langle w^{2}\right\rangle _{f(w)}\right\rangle _{t}-\left\langle \left\langle w\right\rangle _{f(w)}^{2}\right\rangle _{t},
\end{split}
\end{equation}

where in the last step we have defined an induced distribution $f(w)=Z(t)^{-1}\exp\left\{ -\lambda(w)+\sqrt{\hat{q}_{1}}tw\right\} $.
Since the second moments are fixed to 1, we have

\begin{equation}
q=\left\langle \left\langle w\right\rangle _{f(w)}^{2}\right\rangle _{t},\label{appeq:saddle_pt_2_qneq1}
\end{equation}

which gives a nice interpretation of $q$ in terms of the average
overlap of $w$ in the induced distribution $f(w)$.

\subsubsection*{Limit $q\to1$}

We are interested in the critical load $\alpha_{c}$ where the version
space (space of viable weights) shrinks to a single point, i.e., there
exists only one viable solution. Since $q$ measures the typical overlap
between weight vectors in the version space, the uniqueness of the
solution implies $q\to1$ at $\alpha_{c}$. In this limit, the order
parameters $\left\{ \hat{q}_{1},\lambda(w)\right\} $ diverges and
we need to re-derive the saddle point equations Eqn.\ref{appeq:saddle_pt_1_qneq1}
and Eqn.\ref{appeq:saddle_pt_2_qneq1} in terms of the undiverged
order parameters $\left\{ u,r(w)\right\} $:

\begin{equation}
\hat{q}_{1}=\frac{u^{2}}{(1-q)^{2}};\qquad\lambda(w)=\frac{r(w)}{1-q}.
\end{equation}

Now $G_{0}$ becomes

\begin{equation}
G_{0}=\frac{1}{1-q}\left\{ -\frac{1}{2}u^{2}+\int dwq(w)r(w)+(1-q)\left\langle \log Z(t)\right\rangle _{t}\right\} ,
\end{equation}

and 

\begin{equation}
Z(t)\,=\int dw\exp\frac{1}{1-q}\left\{ -r(w)+utw\right\} .
\end{equation}

We can perform a saddle-point approximation for the $w$ integral
in $Z(t)$ at the saddle value $w$ such that $r'(w)=ut$:

\begin{equation}
Z(t)=\exp\left\{ \frac{-r(w)+utw}{1-q}\right\} .\label{appeq:cap_dist_Zt_1}
\end{equation}

Then

\begin{equation}
G_{0}=\frac{1}{1-q}\left\{ -\frac{1}{2}u^{2}+\int dwq(w)r(w)-\left\langle r(w)\right\rangle _{t}+u\left\langle tw\right\rangle \right\} .
\end{equation}

Let's use integration by parts to rewrite

\begin{equation}
\begin{split}\int dwq(w)r(w) & =-\int Q(w)r'(w)dw\\
\left\langle r(w)\right\rangle _{t} & =\int\frac{dt}{\sqrt{2\pi}}e^{-t^{2}/2}r(w)=-\int P(t)r'(w)dw,
\end{split}
\label{appeq:cap_dist_Zt_3}
\end{equation}

where $Q(w)$ is the CDF of the imposed distribution $q(w)$ and $P(t)=\frac{1}{2}\left[1+\text{Erf}(\frac{t}{\sqrt{2}})\right]$
is the normal CDF. 

Now the saddle-point equation

\begin{equation}
0=\frac{\partial G_{0}}{\partial r'(w)}\Rightarrow Q(w)=P(t)\label{appeq:cap_dist_saddle_1}
\end{equation}

determines $w(t)$ implicitly. The $u$ equation gives

\begin{equation}
0=\frac{\partial G_{0}}{\partial u}\Rightarrow u=\left\langle tw\right\rangle _{t}=\left\langle \frac{dw}{dt}\right\rangle _{t}\label{eq:appeq:cap_dist_saddle_2}
\end{equation}

where in the last equality we have used integration by parts. Using
Eqn.\ref{appeq:cap_dist_saddle_1}-\ref{eq:appeq:cap_dist_saddle_2}
$G_{0}$ is simplified to

\begin{equation}
G_{0}=\frac{1}{2(1-q)}\left\langle \frac{dw}{dt}\right\rangle _{t}^{2}.\label{appeq:cap_dist_G0}
\end{equation}

\subsubsection*{The energetic part}

We would like to perform a similar procedure as shown above, to Eqn.\ref{appeq:cap_dist_K_replica}
using the replica-symmetric ansatz. We observe that the effect of
the distribution constraint is entirely captured in $G_{0}$ and therefore
$G_{1}$ is unchanged compared with the standard Gardner calculation
of perceptron capacity. We reproduce the calculation here for completeness.

Under the replica-symmetric ansatz $q_{\alpha\beta}=(1-q)\delta_{\alpha\beta}+q$,
Eqn.\ref{appeq:cap_dist_K_replica} becomes

\begin{equation}
\begin{split}K & =i\sum_{\alpha}x_{\alpha}\rho_{\alpha}-\frac{1-q}{2}\sum_{\alpha}x_{\alpha}^{2}-\frac{q}{2}\left(\sum_{\alpha}x_{\alpha}\right)^{2}\\
 & \HSTeq i\sum_{\alpha}x_{\alpha}\rho_{\alpha}-\frac{1-q}{2}\sum_{\alpha}x_{\alpha}^{2}-it\sqrt{q}\sum_{\alpha}x_{\alpha}.
\end{split}
\end{equation}

where we have again used the Hubbard-Stratonovich transformation to
linearize the quadratic piece. Performing the Gaussian integrals in
$x_{\alpha}$ (define $\alpha=\frac{P}{N}$),

\begin{equation}
nG_{1}=\alpha\log\left[\left\langle \int_{\kappa}^{\infty}\frac{d\rho}{\sqrt{2\pi(1-q)}}\exp\left\{ -\frac{(\rho+t\sqrt{q})^{2}}{2(1-q)}\right\} \right\rangle _{t}^{n}\right].
\end{equation}

At the limit $n\to0$,

\begin{equation}
nG_{1}=\alpha n\left\langle \log\left[\int_{\kappa}^{\infty}\frac{d\rho}{\sqrt{2\pi(1-q)}}\exp\left\{ -\frac{(\rho+t\sqrt{q})^{2}}{2(1-q)}\right\} \right]\right\rangle _{t}.
\end{equation}

Perform the Gaussian integral in $\rho$ and define $\tilde{\kappa}=\frac{\kappa+t\sqrt{q}}{\sqrt{1-q}}$,
we have
\begin{equation}
G_{1}=\alpha\int Dt\log H(\tilde{\kappa}).
\end{equation}

At the limit $q\to1,\alpha\to\alpha_{c}$, $\int_{-\infty}^{\infty}Dt$
is dominated by $\int_{-\kappa}^{\infty}Dt$, and $H(\tilde{\kappa})\to\frac{1}{\sqrt{2\pi}\tilde{\kappa}}e^{-\tilde{\kappa}^{2}/2}$.
The $\mathcal{O}\left(\frac{1}{1-q}\right)$ (leading order) contribution
gives

\begin{equation}
G_{1}=-\frac{1}{2(1-q)}\alpha_{c}\int_{-\kappa}^{\infty}Dt(\kappa+t)^{2}.\label{appeq:cap_dist_G1_1pop}
\end{equation}

Let $G=G_{0}+G_{1}$. As $n\to0$, $\left\langle V^{n}\right\rangle =e^{n\left(NG\right)}\to1+n\left(NG\right)$,
and $\left\langle \log V\right\rangle =\lim_{n\to0}\frac{\left\langle V^{n}\right\rangle -1}{n}=NG$. 

Combining with Eqn.\ref{appeq:cap_dist_G0} (relabel $t\leftrightarrow x$
to distinguish between the two auxiliary Gaussian variables), we have 

\begin{equation}
\left\langle \log V\right\rangle =\frac{N}{2(1-q)}\left[\left\langle \frac{dw}{dx}\right\rangle _{x}^{2}-\alpha_{c}\int_{-\kappa}^{\infty}Dt(\kappa+t)^{2}\right]\label{appeq:cap_dist_logV}
\end{equation}

Capacity $\alpha_{c}$ is reached when Eqn.\ref{appeq:volume_cap_dist}
goes to zero. We arrive at the distribution-constrained capacity 

\begin{equation}
\alpha_{c}(\kappa)=\alpha_{0}(\kappa)\left\langle \frac{dw}{dx}\right\rangle _{x}^{2},\label{appeq:cap_dist_formula}
\end{equation}
where $\alpha_{0}(\kappa)=\left[\int_{-\kappa}^{\infty}Dt(\kappa+t)^{2}\right]^{-1}$
is the unconstrained capacity. 

\subsubsection*{Instructive Examples}

(1) Standard normal distribution $w\sim\mathcal{N}(0,1)$. 

In this case $w=x$ and $\alpha_{c}(\kappa)=\alpha_{0}(\kappa)$.

(2) Normal distribution with nonzero mean $w\sim\mathcal{N}(\mu,\sigma^{2}).$
This is the example discussed in the main text Fig.1.

In this case $w=\mu+\sigma x$ and $\mu^{2}+\sigma^{2}=1$ due to
the second moment constraint Eqn.\ref{appeq:second_moment}. Then
$\alpha_{c}(\kappa)=\sigma^{2}\alpha_{0}(\kappa).$

(3) Lognormal distribution $w\sim\frac{1}{\sqrt{2\pi}w}\exp\left\{ -\frac{(\ln w-\mu)^{2}}{2\sigma^{2}}\right\} .$

In this case $w=e^{\mu+\sigma x}$ where $\mu=-\sigma^{2}.$ $\alpha_{c}(\kappa)=\sigma^{2}e^{-\sigma^{2}}\alpha_{0}(\kappa)$.

\subsubsection*{Geometrical interpretation}

Note that although the Jacobian factor $\left\langle \frac{dw}{dx}\right\rangle _{x}$
takes a simple form, in practice sometimes it might not be the most
convenient form to use. Integrating by parts ($p(x)=\mathcal{N}(0,1)$),

\begin{equation}
\left\langle \frac{dw}{dx}\right\rangle _{x}=\int dxp(x)wx
\end{equation}

Now define $u=P(x)$ so that $du=p(x)dx$ and $w=Q^{-1}(P(x))=Q^{-1}(u)$,
we can express the Jacobian in terms of the CDFs

\begin{equation}
\left\langle \frac{dw}{dx}\right\rangle _{x}=\int_{0}^{1}du\left(Q^{-1}(u)P^{-1}(u)\right)
\end{equation}

Furthermore, 

\begin{equation}
\begin{split}\left\langle \frac{dw}{dx}\right\rangle _{x} & =\frac{1}{2}\left[\int_{0}^{1}du\left(Q^{-1}(u)\right)^{2}+\int_{0}^{1}du\left(P^{-1}(u)\right)^{2}-\int_{0}^{1}du\left(Q^{-1}(u)-P^{-1}(u)\right)^{2}\right]\\
 & =\frac{1}{2}\left[2-W_{2}(P,Q)^{2}\right],
\end{split}
\end{equation}

where we have used second moments equal to $1$ and the definition
of the Wasserstein-$k$ distance in the second equality. Therefore,
we have arrived at the geometric interpretation of the Jacobian term

\begin{equation}
\left\langle \frac{dw}{dx}\right\rangle _{x}=1-\frac{1}{2}W_{2}(P,Q)^{2}.
\end{equation}

\subsubsection{Theory for an arbitrary number of synaptic subpopulations}

\label{app:capacity_M_pop}

In this section, we generalize our theory in the above section to
the set up of a perceptron with $M$ synaptic populations indexed
by $m$, $\boldsymbol{w}^{m}$, such that each $w_{i}^{m}$ satisfies
its own distributions constraints $w_{i}^{m}\sim q_{m}(w^{m})$. We
denote the overall weight vector as $\boldsymbol{w}\equiv\{\boldsymbol{w}^{m}\}_{m=1}^{M}\in\mathbb{R}^{N\times1}$,
where the total number of weights is $N=\sum_{m=1}^{M}N_{m}$. The
replica overlap now becomes $q_{\alpha\beta}=\frac{1}{N}\sum_{m}^{M}\sum_{i}^{N_{m}}w_{i}^{m\alpha}w_{i}^{m\beta}.$
The distribution constraint becomes (see Eqn.\ref{appeq:cap_dist_dist_const_delta}
for the case of $M=1$)

\begin{equation}
\prod_{m}\delta\left(\int dk^{m}\left(\frac{1}{N_{m}}\sum_{i}^{N_{m}}e^{ik^{m}w_{i}^{m\alpha}}-q_{m}(k^{m})\right)\right).
\end{equation}

We introduce $\hat{q}_{\alpha\beta},\lambda_{m}(k)$ to write the
$\delta$-functions into Fourier representations, and use replica-symmetric
ansatz $\hat{q}_{\alpha\beta}=-\frac{i}{2}(\Delta\hat{q}\delta_{\alpha\beta}+\hat{q}_{1})$,
and $q_{\alpha\beta}=(1-q)\delta_{\alpha\beta}+q$ as before. After
similar manipulations that lead to Eqn.\ref{appeq:G0_cap_dist_qneq1},
the entropic part of the free energy becomes ($g_{m}=N_{m}/N$ is
the fraction of weights in $m$-th population)

\begin{equation}
\begin{split}G_{0}= & \frac{1}{2}\Delta\hat{q}-\frac{1}{2}\hat{q}_{1}(1-q)+\sum_{m}g_{m}\int dw^{m}q_{m}(w^{m})\lambda_{m}(w^{m})\\
 & -\frac{1}{2}\Delta\hat{q}\sum_{m}g_{m}\int dw^{m}q_{m}(w^{m})\left(w^{m}\right)^{2}+\sum_{m}g_{m}\left\langle \log Z_{m}(t)\right\rangle _{t},\\
Z_{m}(t)= & \int dw^{m}\exp\left\{ -\lambda_{m}(w^{m})+\sqrt{\hat{q}_{1}}tw^{m}\right\} .
\end{split}
\end{equation}

Now the second moment constraint $0=\partial G_{0}/\partial\Delta\hat{q}$
(Eqn.\ref{appeq:second_moment}) becomes the weighted sum of second
moments from each population:

\begin{equation}
1=\sum_{m}g_{m}\int dw^{m}q_{m}(w^{m})\left(w^{m}\right)^{2}=\sum_{m}g_{m}\left\langle \left(w^{m}\right)^{2}\right\rangle _{q_{m}}.
\end{equation}

We take the $q\to1$ limit as before: 

\begin{equation}
\hat{q}_{1}=\frac{u^{2}}{(1-q)^{2}};\qquad\lambda_{m}(w^{m})=\frac{r_{m}(w^{m})}{1-q}.
\end{equation}

Use saddle-point approximation for $Z_{m}(t)$ and integrate by parts
as in Eqn.\ref{appeq:cap_dist_Zt_1}-\ref{appeq:cap_dist_Zt_3}, the
entropic part becomes

\begin{equation}
G_{0}=\frac{1}{1-q}\left\{ -\frac{1}{2}u^{2}+\sum_{m}g_{m}r'_{m}(w^{m})\left[P(x)-Q_{m}(w^{m})\right]+u\sum_{m}g_{m}\left\langle tw^{m}\right\rangle _{t}\right\} .
\end{equation}

Now the saddle-point equation for order parameters $r'_{m}(w^{m})$
and $u$ gives

\begin{equation}
\begin{split}P(x) & =Q_{m}(w^{m})\\
u & =\sum_{m}g_{m}\left\langle tw^{m}\right\rangle _{t}=\sum_{m}g_{m}\left\langle \frac{dw^{m}}{dt}\right\rangle _{t}.
\end{split}
\end{equation}

Therefore,

\begin{equation}
G_{0}=\frac{1}{2(1-q)}\left[\sum_{m}g_{m}\left\langle \frac{dw^{m}}{dt}\right\rangle _{t}\right]^{2}.
\end{equation}

The energetic part (Eqn.\ref{appeq:cap_dist_Zt_1}) remains unchanged
and thus (relabel $t\leftrightarrow x$) 

\begin{equation}
\alpha_{c}(\kappa)=\alpha_{0}(\kappa)\left[\sum_{m}g_{m}\left\langle \frac{dw^{m}}{dx}\right\rangle _{x}\right]^{2}.
\end{equation}

\subsubsection*{E/I balanced lognormals}

Now we specialize to the biologically realistic E/I balanced lognormal
distributions described in the main text. We are interested the case
with two synaptic populations $m=E,I$ that models the excitatory/inhibitory
synpatic weights of a biological neuron. $w_{i}^{E}\sim\text{\ensuremath{\frac{1}{\sqrt{2\pi}\sigma_{E}w^{E}}\exp\left\{ -\frac{(\ln w^{E}-\mu_{E})^{2}}{2\sigma_{E}^{2}}\right\} }}$
and $w_{i}^{I}\sim\text{\ensuremath{\frac{1}{\sqrt{2\pi}\sigma_{I}w^{I}}\exp\left\{ -\frac{(\ln w^{I}-\mu_{I})^{2}}{2\sigma_{I}^{2}}\right\} }}$.
Let's denote the E/I fractions as $g_{E}=r$ and $g_{I}=1-r$. The
CDF of the lognormals are given by

\begin{equation}
\begin{split}Q_{m}(w^{m})= & H\left[\frac{1}{\sigma_{m}}\left(\mu_{m}-\ln w^{m}\right)\right].\end{split}
\end{equation}

The corresponding inverse CDF is 

\begin{equation}
Q_{m}^{-1}(u)=\exp\left\{ \mu_{m}-\sigma_{m}H^{-1}(u)\right\} .
\end{equation}

The capacity is therefore
\begin{equation}
\begin{split}\alpha_{c} & =\alpha_{0}\left[\sum_{m}g_{m}\int_{0}^{1}duQ_{m}^{-1}(u)P^{-1}(u)\right]^{2}\\
 & =\alpha_{0}\left[r\int_{0}^{1}duH^{-1}(u)\exp\left\{ \mu_{E}-\sigma_{E}H^{-1}(u)\right\} +(1-r)\int duH^{-1}(u)\exp\left\{ \mu_{I}-\sigma_{I}H^{-1}(u)\right\} \right]^{2}.
\end{split}
\end{equation}

This model has five parameters $\left\{ r,\sigma_{E},\sigma_{I},\mu_{E},\mu_{I}\right\} $.
We use values of $r$ reported in experiments (the ratio between of
E. connections found and I. connections found). 

We also have two constraints. The E/I balanced constraint $g_{E}\left\langle w^{E}\right\rangle _{q_{E}}=g_{I}\left\langle w^{I}\right\rangle _{q_{I}}$:

\begin{equation}
re^{\mu_{E}+\frac{1}{2}\sigma_{E}^{2}}=(1-r)e^{\mu_{I}+\frac{1}{2}\sigma_{I}^{2}},
\end{equation}

and the second moment constraint $1=\sum_{m}g_{m}\left\langle \left(w^{m}\right)^{2}\right\rangle _{q_{m}}$:

\begin{equation}
1=re^{2(\mu_{E}+\sigma_{E}^{2})}+(1-r)e^{2(\mu_{I}+\sigma_{I}^{2})}.
\end{equation}

Therefore there are two free parameters left and we choose to express
$\mu_{E}$ and $\mu_{I}$ in terms of the rest:

\begin{equation}
\begin{split}\mu_{I}= & -\frac{1}{2}\sigma_{I}^{2}-\ln(1-r)-\frac{1}{2}\ln\left[\frac{e^{\sigma_{I}^{2}}}{1-r}+\frac{e^{\sigma_{E}^{2}}}{r}\right]\\
\mu_{E}= & -\frac{1}{2}\sigma_{E}^{2}-\ln r-\frac{1}{2}\ln\left[\frac{e^{\sigma_{I}^{2}}}{1-r}+\frac{e^{\sigma_{E}^{2}}}{r}\right].
\end{split}
\end{equation}

The parameter landscape is plotted against the two free parameters
$\sigma_{E}$ and $\sigma_{I}$. Here we report comparisons across
different experiments \cite{levy2012spatial,avermann2012microcircuits,holmgren2003pyramidal,molnar2008complex,thomson2002synaptic,yang2013development}
similar to main text Fig.4 (Fig.4 (a) is included here for reference).
Note that despite the apparently different shape of distributions,
all the experimentally measured parameter values are within the first
quantile of the optimal values predicted by our theory.

\begin{figure}
\centering{}\includegraphics[scale=0.12]{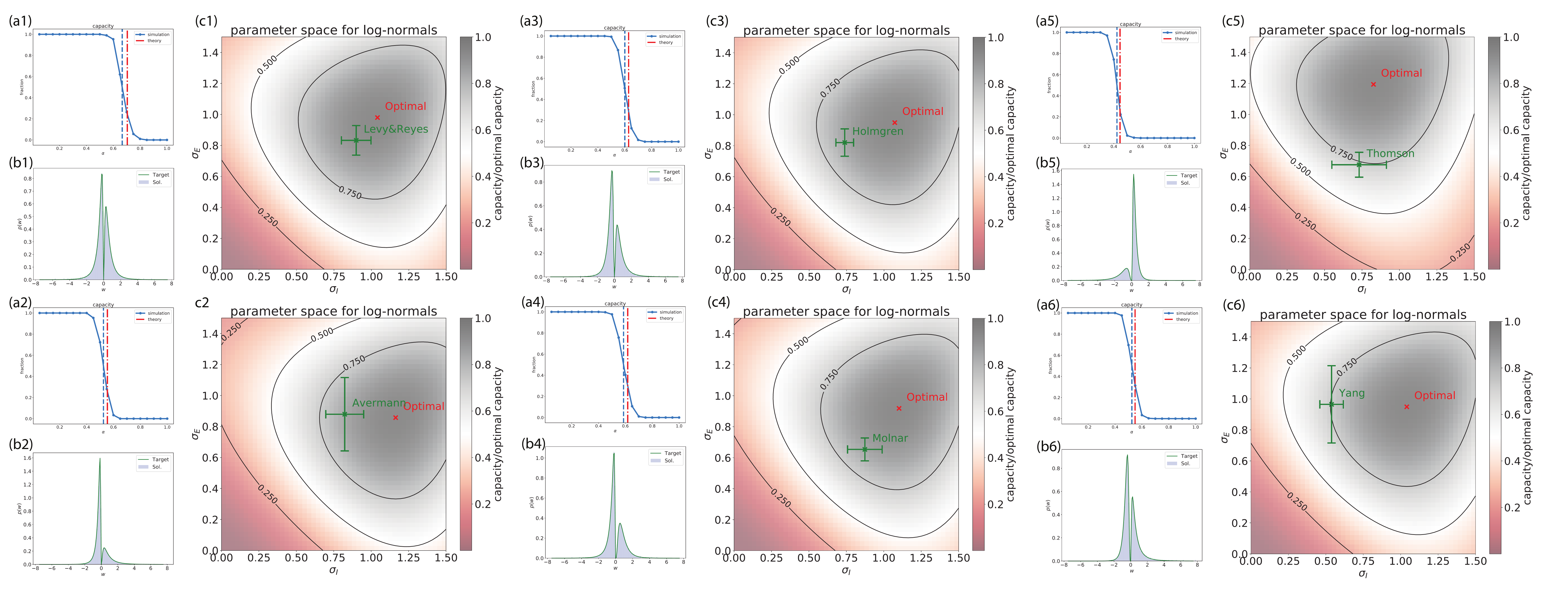}\caption{\label{appfig:Experimental_panel} Additional parameter landscape
for the biologically-realistic distribution. (a)-(b) (theory from
main text Eqn.10 and simulations from DisCo-SGD): (a) Determination
of capacity; (b) Example of weight distribution obtained in simulation.
(c) Capacity (normalized by the optimal value in the landscape) as
a function of the lognormal parameters $\sigma_{E}$ and $\sigma_{I}$.
Experimental value is shown in green with error bars, and optimal
capacity is shown in red. }
\end{figure}

\subsubsection{Capacity for biased inputs and sparse label}

\label{app:capacity_biased_sparse}

In this section, we generalized our theory in Section \ref{app:capacity_dist_const}
to the set up of nonzero-mean input patterns $\boldsymbol{\xi}^{\mu}$
and sparse labels $\zeta^{\mu}$:

\begin{equation}
\begin{split}p(\xi_{i}^{\mu})= & \mathcal{N}(m,1-m^{2})\\
p(\zeta^{\mu})= & f\delta(\zeta^{\mu}-1)+(1-f)\delta(\zeta^{\mu}+1).
\end{split}
\end{equation}

In this case, we need to include a bias in the perceptron $\hat{\zeta}^{\mu}=\text{sgn}(\frac{\boldsymbol{w}\cdot\boldsymbol{\xi}^{\mu}}{\left\Vert \boldsymbol{w}\right\Vert }-b)$
to be able to correctly classify patterns in general. 

Note that $m=0$ and $f=1/2$ reduces to the case in Section \ref{app:capacity_dist_const}.
We observe due to the multiplicative relation between the Jacobian
term and the original Gardner capacity in Eqn.\ref{appeq:cap_dist_formula},
entropic effects (such as distribution constraints and sign-constraints)
factors with the energetic effects (such as the nonzero mean inputs
and sparse labels), and they don't interfere with each other. Therefore,
the calculations for nonzero mean inputs and sparse labels are identical
with the original Gardner case. Here we only reproduce the calculation
for completeness. Readers already familiar with this calculation should
skip this part. 

The analog of Eqn.\ref{appeq:cap_dist_energy_x_z} reads (define the
local fields as $h_{i}^{\mu}=\sum_{\alpha}x_{\alpha}^{\mu}w_{i}^{\alpha}$)

\begin{equation}
\begin{split}\prod_{\mu\alpha}\left\langle e^{-\frac{i}{\sqrt{N}}x_{\alpha}^{\mu}\zeta^{\mu}\boldsymbol{\xi}^{\mu}\cdot\boldsymbol{w}^{\alpha}}\right\rangle _{\xi\zeta}= & \prod_{\mu i}\left\langle \exp\left\{ -\frac{i}{\sqrt{N}}\zeta^{\mu}\xi_{i}^{\mu}h_{i}^{\mu}\right\} \right\rangle _{\xi\zeta}\\
= & \prod_{\mu i}\left\langle \exp\left\{ -\frac{im}{\sqrt{N}}\zeta^{\mu}h_{i}^{\mu}-\frac{1}{2N}(1-m^{2})\left(h_{i}^{\mu}\right)^{2}\right\} \right\rangle _{\zeta}\\
= & \prod_{\mu}\left\langle \exp\left\{ -im\zeta^{\mu}\sum_{\alpha}x_{\alpha}^{\mu}M_{\alpha}-\frac{1-m^{2}}{2}\sum_{\alpha\beta}x_{\alpha}^{\mu}x_{\beta}^{\mu}q_{\alpha\beta}\right\} \right\rangle _{\zeta},
\end{split}
\end{equation}

where in the second equality we have carried out the Gaussian integral
in $\boldsymbol{\xi}^{\mu}$ and in the third equality we introduced
the order parameters 

\begin{equation}
q_{\alpha\beta}=\frac{1}{N}\sum_{i}w_{i}^{\alpha}w_{i}^{\beta},\qquad M_{\alpha}=\frac{1}{\sqrt{N}}\sum_{i}w_{i}^{\alpha}.
\end{equation}

Now the full energetic term becomes

$\begin{aligned}\left\langle \Theta\left(\frac{1}{\sqrt{N}}\zeta^{\mu}\boldsymbol{\xi}^{\mu}\cdot\boldsymbol{w}^{\alpha}-b\zeta^{\mu}-\kappa\right)\right\rangle _{\xi\zeta}\qquad\qquad\qquad\qquad\\
=\prod_{\mu}\left\langle \int_{\kappa+b\zeta^{\mu}}^{\infty}\frac{d\lambda_{\alpha}^{\mu}}{2\pi}\int dx_{\alpha}^{\mu}\exp\left\{ -im\zeta^{\mu}\sum_{\alpha}x_{\alpha}^{\mu}M_{\alpha}-\frac{1-m^{2}}{2}\sum_{\alpha\beta}x_{\alpha}^{\mu}x_{\beta}^{\mu}q_{\alpha\beta}\right\} \right\rangle _{\zeta}\\
=f\prod_{\mu}\int_{\kappa+b}^{\infty}\frac{d\lambda_{\alpha}^{\mu}}{2\pi}\int dx_{\alpha}^{\mu}\exp\left\{ i\sum_{\alpha}x_{\alpha}^{\mu}\left(\lambda_{\alpha}^{\mu}-mM_{\alpha}\right)-\frac{1-m^{2}}{2}\sum_{\alpha\beta}x_{\alpha}^{\mu}x_{\beta}^{\mu}q_{\alpha\beta}\right\} \\
+(1-f)\prod_{\mu}\int_{\kappa-b}^{\infty}\frac{d\lambda_{\alpha}^{\mu}}{2\pi}\int dx_{\alpha}^{\mu}\exp\left\{ i\sum_{\alpha}x_{\alpha}^{\mu}\left(\lambda_{\alpha}^{\mu}+mM_{\alpha}\right)-\frac{1-m^{2}}{2}\sum_{\alpha\beta}x_{\alpha}^{\mu}x_{\beta}^{\mu}q_{\alpha\beta}\right\} \\
=f\prod_{\mu}\int_{\frac{\kappa+b-mM_{\alpha}}{\sqrt{1-m^{2}}}}^{\infty}\frac{d\lambda_{\alpha}^{\mu}}{2\pi}\int dx_{\alpha}^{\mu}\exp\left\{ i\sum_{\alpha}x_{\alpha}^{\mu}\lambda_{\alpha}^{\mu}-\frac{1}{2}\sum_{\alpha\beta}x_{\alpha}^{\mu}x_{\beta}^{\mu}q_{\alpha\beta}\right\} \\
+(1-f)\prod_{\mu}\int_{\frac{\kappa-b+mM_{\alpha}}{\sqrt{1-m^{2}}}}^{\infty}\frac{d\lambda_{\alpha}^{\mu}}{2\pi}\int dx_{\alpha}^{\mu}\exp\left\{ i\sum_{\alpha}x_{\alpha}^{\mu}\lambda_{\alpha}^{\mu}-\frac{1}{2}\sum_{\alpha\beta}x_{\alpha}^{\mu}x_{\beta}^{\mu}q_{\alpha\beta}\right\} .
\end{aligned}
$

Now $G_{1}$ becomes 

\begin{equation}
\begin{split}G_{1}= & \frac{1}{1-q}\left\{ f\int_{\frac{\kappa-b+mM}{\sqrt{1-m^{2}}}}^{\infty}Dt\left(t+\frac{\kappa+b-mM}{\sqrt{1-m^{2}}}\right)^{2}+(1-f)\int_{\frac{-\kappa-b-mM}{\sqrt{1-m^{2}}}}^{\infty}Dt\left(t+\frac{\kappa-b+mM}{\sqrt{1-m^{2}}}\right)^{2}\right\} \end{split}
.
\end{equation}

Note that the hat-variables $\hat{M}_{\alpha}$ conjugated with $M_{\alpha}$
are in subleading order to $\hat{q}_{\alpha\beta}$ in the thermodynamic
limit, and therefore $G_{0}$ is unchanged. Let $v=M-b/m$, we have
now the capacity

\begin{equation}
\begin{split}\alpha_{c}(\kappa)= & \left\langle \frac{dw}{dx}\right\rangle _{x}^{2}\left[f\int_{\frac{-\kappa+mv}{\sqrt{1-m^{2}}}}^{\infty}Dt\left(t+\frac{\kappa-mv}{\sqrt{1-m^{2}}}\right)^{2}+(1-f)\int_{\frac{-\kappa-mv}{\sqrt{1-m^{2}}}}^{\infty}Dt\left(t+\frac{\kappa+mv}{\sqrt{1-m^{2}}}\right)^{2}\right]^{-1}\end{split}
,
\end{equation}

where the order parameter $v$ needs to be determined from the saddle-point
equation

\begin{equation}
f\int_{\frac{-\kappa+mv}{\sqrt{1-m^{2}}}}^{\infty}Dt\left(t+\frac{\kappa-mv}{\sqrt{1-m^{2}}}\right)=(1-f)\int_{\frac{-\kappa-mv}{\sqrt{1-m^{2}}}}^{\infty}Dt\left(t+\frac{\kappa+mv}{\sqrt{1-m^{2}}}\right).
\end{equation}

\begin{figure}
\centering{}\includegraphics[scale=0.36]{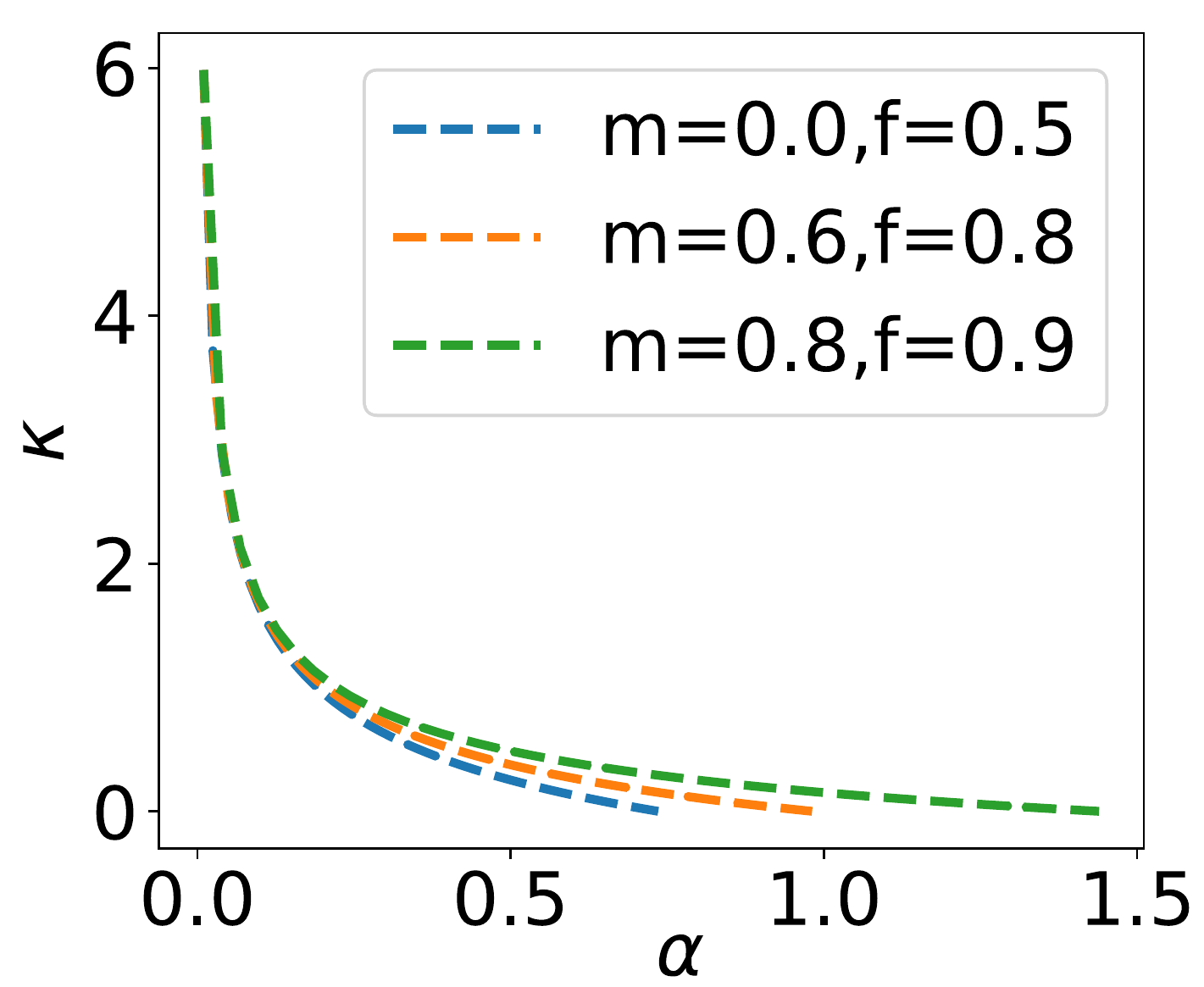}\caption{\label{appfig:nonzero_mean_inputs}$\alpha_{c}(\kappa)$ for different
values of input mean $m$ and label sparsity $f$. Note that the blue
curve corresponds to the vanilla case shown in main text Fig.4(c).}
\end{figure}

In Fig.\ref{appfig:nonzero_mean_inputs} we numerically solve $\alpha_{c}(\kappa)$
for different values of $m$ and $f$. 

\subsection{Optimal transport theory}

\label{app:Optimal-transport-theory}

In recent years, Wasserstein distances has found diverse applications
in fields ranging from machine learning \cite{arjovsky2017wasserstein,frogner2015learning,montavon2016wasserstein}
to geophysics \cite{engquist2013application,engquist2016optimal,chen2018quadratic,metivier2016measuring,metivier2016optimal}.
In optimal transport theory, the Wasserstein-$k$ distance arise as
the minimal cost one needs to pay in transporting one probability
distribution to another, when the moving cost between probability
masses are measured by the $L_{k}$ norm \cite{villani2009optimal}.
When one equips the probability density manifold with the Wasserstein-$2$
distance as metric, it becomes the Wasserstein space, a Riemannian
manifold of real-valued distributions with a constant nonnegative
sectional curvature \cite{lott2006some,figalli2011optimal,chen2020optimal}.
Note that in our statistical mechanical theory main text Eqn.3-5,
the Wasserstein-$2$ distance naturally arises in the mean-field limit
without assuming any a priori transportation cost.

Here we briefly review the theory of optimal transport. Intuitively,
optimal transport concerns the problem of finding the shortest path
of morphing one distribution into another. In the following, we will
use the \textit{Monge} formulation \cite{thorpe2019introduction,ambrosio2013user}.

Given probability distributions $P$ and $Q$ with supports $X$ and
$Y$, we say that $T:X\to Y$ is a transport map from $P$ to $Q$
if the \textit{push-forward }of $P$ through $T,$ $T_{\#}P$, equals
$Q$:

\begin{equation}
Q=T_{\#}P\equiv P(T^{-1}(Y)).\label{eq:pushforward}
\end{equation}

Eqn.\ref{eq:pushforward} can be understood as moving probability
masses $x\in X$ from distribution $P$ to $y\in Y$ according to
transportation map $T$, such that upon completion the distribution
over $Y$ becomes $Q$.

We are interested in finding a transportation plan that minimizes
the transportation cost as measured by some distance function $d:X\times Y\to\mathbb{R}$
:

\begin{equation}
C(T;d)=\int_{X}d(T(x),x)p(x)dx\qquad\text{s.t.}\;T_{\#}P=Q.\label{eq:transportation_cost}
\end{equation}

The transportation plan that minimizes Eqn.\ref{eq:transportation_cost}
is called the optimal transport plan $T^{*}=\text{argmin}_{T}C(T;d)$.
When the distance function $d$ is chosen to be the $L_{k}$ norm,
the minimal cost becomes the Wasserstein-$k$ distance:

\begin{equation}
W_{k}(P,Q)=\inf_{T}C(T;L_{k})|_{T_{\#}P=Q}.\label{eq:Wass-p}
\end{equation}

\begin{figure}
\centering{}\includegraphics[scale=0.18]{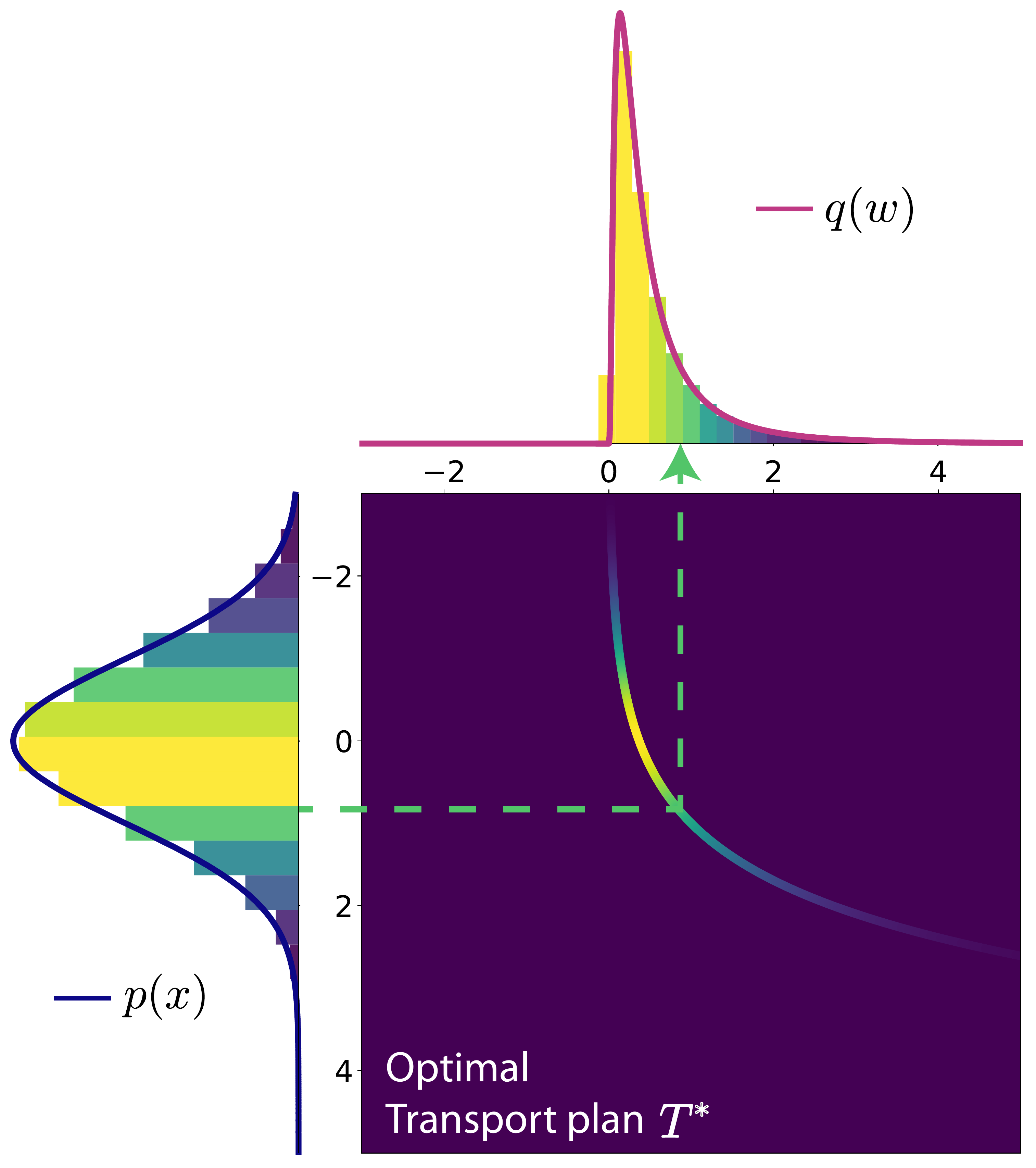}\caption{\label{appfig:optimal_transport_lognormal}An example optimal transport
plan from standard normal, $p(x)$, to a lognormal distribution $q(w)$.
The optimal transport plan $T^{*}$ is plotted in between the distributions.
$T^{*}$ moves $p(x)$ units of probability mass $x$ to location
$w$, as indicated by the dashed line, and the colors are chosen to
reflect the amount of probability mass to be transported.}
\end{figure}

In $1$-dimension, the Wasserstein-$k$ distance has a closed form
given by main text Eqn.6, and the optimal transport map has an explicit
formula in terms of the CDFs: $T^{*}=Q^{-1}\circ P$. An example of
the optimal transport map and how it moves probability masses between
distributions is given in Fig.\ref{appfig:optimal_transport_lognormal}
for transport between $p(w)=\mathcal{N}(0,1)$ and $q(w)=\frac{1}{\sqrt{2\pi}\sigma w}\exp\left\{ \frac{(\ln w-\mu)^{2}}{2\sigma^{2}}\right\} .$
Note that in this case, the optimal transport plan is simply $T^{*}(x)=e^{\mu+\sigma x}$.

Now consider the manifold $\mathcal{M}$ of real-valued probability
distributions, where points on this manifold are probability measures
that admits a probability density function. When endowed with the
$W_{k}$ metric, $(\mathcal{M},W_{k})$ becomes a metric space and
is in particular a geodesic space \cite{thorpe2019introduction,ambrosio2013user}.
We can explicitly construct the geodesics connecting points on $\mathcal{M}$.
We parameterize the geodesic by the \textit{geodesic time} $\tau\in[0,1].$
Then given $T^{*}$ an optimal transport plan, the intermediate probability
distributions along the geodesic take the following form \cite{thorpe2019introduction}:

\begin{equation}
P_{\tau}=\left((1-\tau)\text{Id}+\tau T^{*}\right)_{\#}P\label{eq:geodesic}
\end{equation}

where $\text{Id}$ is the identity map and $P_{\tau}$ is a constant
speed geodesic connecting $P_{\tau=0}=P$ and $P_{\tau=1}=Q$.

For the discrete case, we can describe the sample $\left\{ w_{i}^{\tau}\right\} $
from $P_{\tau}$ in a simple manner in terms of the samples $\left\{ w_{i}\right\} $
drawn from $P$ and $\left\{ \hat{w}_{i}\right\} $ drawn from $Q$.
We can arrange the samples in the ascending order, or equivalently,
forming their order statistics $\left\{ x_{(i)}:x_{(1)}\leq...\leq x_{(N)}\right\} $,
which can be thought of as atoms in a discrete measure. Then in terms
of the order statistics, 

\begin{equation}
w_{(i)}^{\tau}=(1-\tau)w_{(i)}+\tau\hat{w}_{(i)}
\end{equation}

Upon infinitetesimal change in the geodesic time, $\tau\to\tau+\delta\tau$,
the geodesic flow becomes

\begin{equation}
w_{(i)}^{\tau+\delta\tau}=w_{(i)}^{\tau}+\delta\tau\left(\hat{w}_{(i)}-w_{(i)}\right)
\end{equation}

Specializing to the case discussed in main text Section 3, $w_{(i)}=w_{(i)}^{\tau=0}$
is the initialization for the perceptron weight and therefore just
a constant, we can promoted it $w_{(i)}\to w_{(i)}^{\tau}$ to fix
the overall scale in the perceptron weight, then we arrive at main
text Eqn.9.

\subsection{Generalization supplemental materials}

\label{app:generalization}

\subsubsection{Replica calculation of generalization with sign-constraint}

\label{app:generalization_sign_const}

In this section, we calculate the sign-constraint teacher-student
setup. To ease notation, let's denote the teacher perceptron $\boldsymbol{w_{t}}\equiv\boldsymbol{w}^{0}$
and the (replicated) student perceptron $\boldsymbol{w}_{s}^{a}\equiv\boldsymbol{w}^{a}.$
Given random inputs $\boldsymbol{\xi}^{\mu}$ with $p(\xi_{i}^{\mu})=\mathcal{N}(0,1)$,
we generate labels by $\zeta^{\mu}=\text{sgn}(\boldsymbol{w}^{0}\cdot\boldsymbol{\xi}^{\mu}/||\boldsymbol{w}^{0}||+\eta^{\mu})$,
where $\eta^{\mu}$ is input noise and $\eta^{\mu}\sim\mathcal{N}(0,\sigma^{2})$.
Let's denote the signs of the teacher perceptron as $s_{i}=\text{sgn}(w_{i}^{0}).$
The student perceptron's weights are constrained to have the same
sign as that of the teacher's, so we insert $\Theta(s_{i}w_{i}^{a})$
in the Gardner volume to enforce this constraint (we leave out the
denominator part of $V$ as it does not depend on data and is an overall
constant): 

\begin{equation}
\left\langle V^{n}\right\rangle _{\xi\eta w^{0}}=\prod_{\alpha=1}^{n}\left\langle \int_{-\infty}^{\infty}\frac{d\boldsymbol{w}^{a}}{\sqrt{2\pi}}\prod_{\mu=1}^{p}\Theta\left(\text{sgn}\left(\frac{\boldsymbol{w^{0}}\cdot\boldsymbol{\xi}^{\mu}}{||\boldsymbol{w^{0}}||}+\eta^{\mu}\right)\frac{\boldsymbol{w^{a}}\cdot\boldsymbol{\xi}^{\mu}}{||\boldsymbol{w^{a}}||}-\kappa\right)\prod_{i}^{N}\Theta(s_{i}w_{i}^{a})\right\rangle _{\xi\eta w^{0}}.\label{sc_original}
\end{equation}

We observe that upon redefining $s_{i}w_{i}^{a}\to w_{i}^{a},s_{i}\xi_{i}^{\mu}\to\xi_{i}^{\mu}$,
we can absorb the sign-constraints into the integration range of $w$
from $[-\infty,+\infty]$ to $[0,\infty]$:

\begin{equation}
\left\langle V^{n}\right\rangle _{\xi\eta w^{0}}=\prod_{\alpha=1}^{n}\left\langle \int_{0}^{\infty}\frac{d\boldsymbol{w}^{a}}{\sqrt{2\pi}}\prod_{\mu=1}^{p}\Theta\left(\text{sgn}\left(\frac{\boldsymbol{w^{0}}\cdot\boldsymbol{\xi}^{\mu}}{||\boldsymbol{w^{0}}||}+\eta^{\mu}\right)\frac{\boldsymbol{w^{a}}\cdot\boldsymbol{\xi}^{\mu}}{||\boldsymbol{w^{a}}||}-\kappa\right)\right\rangle _{\xi\eta w^{0}}.\label{sc_vol}
\end{equation}

Therefore, sign constraint amounts to restricting all the weights
to be positive. In the following, we denote $\int_{0}^{\infty}$as
$\int_{c}$. 

Let's define the local fields as
\begin{equation}
h_{\mu}^{a}=\frac{\boldsymbol{w^{a}}\cdot\boldsymbol{\xi}^{\mu}}{\sqrt{N}};\qquad h_{\mu}^{0}=\frac{\boldsymbol{w^{0}}\cdot\boldsymbol{\xi}^{\mu}}{\sqrt{N}}+\eta^{\mu}
\end{equation}

We leave the average over teacher $w^{0}$ to the very end.

\begin{equation}
\begin{split}\left\langle V^{n}\right\rangle _{\xi\eta} & =\prod_{\mu a}\int_{c}\frac{d\boldsymbol{w}^{a}}{\sqrt{2\pi}}\int dh_{\mu}^{a}\Theta\bigg(\text{sgn}(h_{\mu}^{0})h_{\mu}^{a}-\kappa\bigg)\left\langle \delta\left(h_{\mu}^{a}-\frac{\boldsymbol{w^{a}}\cdot\boldsymbol{\xi}^{\mu}}{\sqrt{N}}\right)\right\rangle _{\xi\eta}\\
 & =\int_{c}(\prod_{a=1}^{n}\frac{d\boldsymbol{w}^{a}}{\sqrt{2\pi}})\int\prod_{\mu a}\frac{dh_{\mu}^{a}d\hat{h}_{\mu}^{a}}{2\pi}\int\prod_{\mu}\frac{dh_{\mu}^{0}d\hat{h}_{\mu}^{0}}{2\pi}\prod_{\mu a}\Theta\bigg(\text{sgn}(h_{\mu}^{0})h_{\mu}^{a}-\kappa\bigg)\\
 & \times\bigg\langle\exp\bigg\{\sum_{\mu a}\bigg(i\hat{h}_{\mu}^{a}h_{\mu}^{a}-i\hat{h}_{\mu}^{a}\frac{\boldsymbol{w^{a}}\cdot\boldsymbol{\xi}^{\mu}}{\sqrt{N}}\bigg)+\sum_{\mu}\bigg(i\hat{h}_{\mu}^{0}h_{\mu}^{0}-i\hat{h}_{\mu}^{0}\frac{\boldsymbol{w^{0}}\cdot\boldsymbol{\xi}^{\mu}}{\sqrt{N}}-i\hat{h}_{\mu}^{0}\eta^{\mu}\bigg)\bigg\}\bigg\rangle_{\xi\eta}\\
 & =\int_{c}(\prod_{a=1}^{n}\frac{d\boldsymbol{w}^{a}}{\sqrt{2\pi}})\int\prod_{\mu a}\frac{dh_{\mu}^{a}d\hat{h}_{\mu}^{a}}{2\pi}\int\prod_{\mu}\frac{dh_{\mu}^{0}d\hat{h}_{\mu}^{0}}{2\pi}\prod_{\mu a}\Theta\bigg(\text{sgn}(h_{\mu}^{0})h_{\mu}^{a}-\kappa\bigg)\\
 & \times\exp\left\{ \sum_{\mu a}i\hat{h_{\mu}^{a}}h_{\mu}^{a}+\sum_{\mu}i\hat{h}_{\mu}^{0}h_{\mu}^{0}\right\} \\
 & \times\prod_{\mu}\exp\left\{ -\frac{1}{2N}\left[\sum_{a,b}\hat{h}_{\mu}^{a}\hat{h}_{\mu}^{b}\sum_{i}w_{i}^{a}w_{i}^{b}+N\left(\hat{h}_{\mu}^{0}\right)^{2}+2\sum_{a}\hat{h}_{\mu}^{a}\hat{h}_{\mu}^{0}\sum_{i}w_{i}^{a}w_{i}^{0}\right]\right\} ,
\end{split}
\end{equation}

where in the last step we perform the average over noise $\eta^{\mu}\sim\mathcal{N}(0,\sigma^{2})$
and patterns $p(\xi_{i}^{\mu})=\mathcal{N}(0,1)$, and make use of
the normalization conditions $\sum_{i}(w_{i}^{0})^{2}=N$ and $\sum_{i}(w_{i}^{a})^{2}=N$.

Now let's define order parameters
\begin{equation}
q_{ab}=\frac{1}{N}\sum_{i}w_{i}^{a}w_{i}^{b},\qquad R_{a}=\frac{1}{N}\sum_{i}w_{i}^{a}w_{i}^{0}.
\end{equation}

We introduce conjugate variables $\hat{q}_{ab}$ and $\hat{R}_{a}$
to write the $\delta$-functions into its Fourier representations,
and after some algebraic manipulations we can bring the Gardner volume
into the following form ($\alpha\equiv p/N$):

\begin{equation}
\begin{split}\langle\langle V^{n}\rangle\rangle_{\xi,z} & =\int(\prod_{a}d\hat{q}_{1}^{a})(\prod_{ab}dq^{ab}d\hat{q}^{ab})(\prod_{a}dR^{a}d\hat{R}^{a})e^{nNG}\end{split}
,
\end{equation}

where ($\bar{h}_{\mu}^{0}=\gamma h_{\mu}^{0};\quad\gamma=1/\sqrt{1+\sigma^{2}}$)

\begin{equation}
\begin{split}nG= & nG_{0}+\alpha nG_{E}\\
nG_{0}= & -\frac{1}{2}\sum_{ab}\hat{q}^{ab}q^{ab}-\sum_{a}\hat{R}^{a}R^{a}+n\left\langle \ln Z\right\rangle _{w^{0}},\\
Z= & \int_{c}\left(\prod_{a}\frac{dw_{i}^{a}}{\sqrt{2\pi}}\right)\exp\bigg\{\frac{1}{2}\sum_{a}\hat{q}_{1}^{a}(w_{i}^{a})^{2}+\frac{1}{2}\sum_{a\neq b}\hat{q}^{ab}w_{i}^{a}w_{i}^{b}+\sum_{a}\hat{R}^{a}w_{i}^{a}w_{i}^{0}\bigg\},\\
nG_{1}= & \ln\int\prod_{a}\frac{d\hat{h}^{a}dh^{a}}{2\pi}\int D\bar{h}^{0}\prod_{a}\Theta\bigg(\text{sgn}(\frac{\bar{h}^{0}}{\gamma})h^{a}-\kappa\bigg)\\
 & \times\exp\bigg\{ i\sum_{a}\hat{h}^{a}h^{a}-i\gamma\bar{h}^{0}\sum_{a}h^{a}R^{a}-\frac{1}{2}\sum_{a}(\hat{h}^{a})^{2}[1-(\gamma R^{a})^{2}]-\frac{1}{2}\sum_{a\neq b}\hat{h}^{a}\hat{h}^{b}(q^{ab}-\gamma^{2}R^{a}R^{b})\bigg\}.
\end{split}
\end{equation}

The energetic part $G_{1}$ is the same as the unconstrained case
in \cite{seung1992statistical,engel2001statistical}. After standard
manipulations, we have
\begin{equation}
G_{1}=2\int DtH\bigg(-\frac{\gamma Rt}{\sqrt{q-\gamma^{2}R^{2}}}\bigg)\ln H\bigg(\frac{\kappa-\sqrt{q}t}{\sqrt{1-q}}\bigg).
\end{equation}

\subsubsection*{Entropic part}

In this subsection, we perform the integrals in the entropic part,
and we will see novel terms coming from the constraint on the student's
integration range.

We start by assuming a replica-symmetric solution for the auxiliary
variables introduced in the Fourier decomposition of the $\delta$-functions,
\begin{equation}
\hat{R}^{a}=\hat{R};\qquad\hat{q}^{ab}=\hat{q}+(\hat{q}_{1}-\hat{q})\delta_{ab};\qquad\hat{q}_{1}^{a}=\hat{q}_{1};\qquad m_{i}^{a}=m_{i};\qquad\hat{m}_{i}^{a}=\hat{m}_{i},
\end{equation}

and $q_{ab}=(1-q)\delta_{ab}+q.$

Then the entropic part, 
\begin{equation}
\begin{split}Z & =\int\left(\prod_{a}\frac{dw_{i}^{a}}{\sqrt{2\pi}}\right)\exp\bigg\{\frac{1}{2}(\hat{q}_{1}-\hat{q})\sum_{a}(w_{i}^{a})^{2}+\hat{R}w_{i}^{0}\sum_{a}w_{i}^{a}+\frac{1}{2}\hat{q}(\sum_{a}w_{i}^{a})^{2}\bigg\}\\
 & \HSTeq\int Dt\int_{c}(\prod_{a}\frac{dw_{i}^{a}}{\sqrt{2\pi}})\exp\bigg\{\frac{1}{2}(\hat{q}_{1}-\hat{q})\sum_{a}(w_{i}^{a})^{2}+(\hat{R}w_{i}^{0}+t\sqrt{\hat{q}})\sum_{a}w_{i}^{a}\bigg\},
\end{split}
\end{equation}

where we have introduced Gaussian variable $t$ to linearize quadratic
term as usual. Now the integral becomes $n$ identical copies and
we can drop the replica index $a$,
\begin{equation}
G_{0}=-\frac{1}{2}\hat{q}_{1}+\frac{1}{2}\hat{q}q-\hat{R}R+\left\langle \ln Z\right\rangle _{t,w^{0}}.
\end{equation}

We can bring the log term into the form of an induced distribution
$f(w)$,

\begin{equation}
\begin{split}Z= & \int_{0}^{\infty}\frac{dw}{\sqrt{2\pi}}\exp\left[-f(w)\right]\\
f(w)= & \frac{1}{2}(\hat{q}-\hat{q}_{1})w^{2}-(\hat{R}w^{0}+t\sqrt{\hat{q}})w
\end{split}
.
\end{equation}

Under saddle-point approximation, we obtain a set of mean field self-consistency
equations for the order parameters: 
\begin{equation}
\begin{split}0=\frac{\partial G_{0}}{\partial\hat{q}_{1}} & \Rightarrow1=\left\langle \left\langle w^{2}\right\rangle _{f}\right\rangle _{t,w^{0}}\\
0=\frac{\partial G_{0}}{\partial\hat{R}} & \Rightarrow R=\left\langle w^{0}\left\langle w\right\rangle _{f}\right\rangle _{t,w^{0}}\\
0=\frac{\partial G_{0}}{\partial\hat{q}} & \Rightarrow q=\left\langle \left\langle w\right\rangle _{f}^{2}\right\rangle _{t,w^{0}}
\end{split}
,\label{appeq:sc_saddle_1-3_qneq1}
\end{equation}

\begin{equation}
\begin{split}0=\frac{\partial G_{1}}{\partial q} & \Rightarrow\hat{q}=-2\alpha\partial_{q}G_{1}\\
0=\frac{\partial G_{1}}{\partial R} & \Rightarrow\hat{R}=\alpha\partial_{R}G_{1}
\end{split}
.\label{appeq:sc_saddle_4-5_qneq1}
\end{equation}

\subsubsection*{$q\to1$ limit}

In this limit the order parameter diverges, and we define the new
set of undiverged order parameters as

\begin{equation}
\hat{R}=\frac{\tilde{R}}{1-q};\qquad\hat{q}=\frac{\tilde{q}^{2}}{(1-q)^{2}};\qquad\hat{q}-\hat{q}_{1}=\frac{\Delta}{1-q}.
\end{equation}

Then 

\begin{equation}
\begin{split}f(w)= & \frac{1}{1-q}\left[\frac{1}{2}\Delta w^{2}-(\tilde{R}w^{0}+t\tilde{q})w\right]\\
= & \frac{1}{1-q}\left[\frac{1}{2}\Delta\left(w-\frac{1}{\Delta}(\tilde{R}w^{0}+t\tilde{q})\right)^{2}-\frac{1}{2\Delta}(\tilde{R}w^{0}+t\tilde{q})^{2}\right].
\end{split}
\end{equation}

Then $\langle w\rangle_{f}=\frac{1}{\Delta}\left(\tilde{R}w^{0}+t\tilde{q}\right)$,
and the integral over the auxiliary variable is dominated by values
of $t$ such that $\tilde{R}w^{0}+t\tilde{q}>0$. In the following,
we denote $\left\langle \left[g(t)\right]_{+}\right\rangle _{t}$
as integrating over range of $t$ such that $g(t)>0$. Then the self-consistency
equations Eqn.\ref{appeq:sc_saddle_1-3_qneq1} take a compact form
(after rescaling order parameters $\tilde{R}\rightarrow\tilde{R}\Delta$
, $\tilde{q}\rightarrow\tilde{q}\Delta$)

\begin{equation}
\begin{split}1= & \frac{1}{\Delta}\left\langle \Theta(\tilde{R}w^{0}+t\tilde{q})\right\rangle _{t,w^{0}}\\
1= & \left\langle \left[\tilde{R}w^{0}+t\tilde{q}\right]_{+}^{2}\right\rangle _{t,w^{0}}\\
R= & \left\langle w^{0}\left[\tilde{R}w^{0}+t\tilde{q}\right]_{+}\right\rangle _{t,w^{0}}
\end{split}
,\label{appeq:sc_saddle_1-3_q=00003D1}
\end{equation}

Eqn.\ref{appeq:sc_saddle_1-3_qneq1} becomes ($\tilde{\kappa}=\kappa/\sqrt{1-\gamma^{2}R^{2}}$)

\begin{equation}
\begin{split}\tilde{R}\Delta & =\frac{\alpha\gamma}{\sqrt{2\pi}}\sqrt{1-\gamma^{2}R^{2}}\int_{-\tilde{\kappa}}^{\infty}Dt\bigg(\tilde{\kappa}+t\bigg)\\
\frac{\Delta}{2}\left(2-\tilde{q}^{2}\Delta-2\tilde{R}R\right)= & \alpha\int_{-\infty}^{\kappa}DtH\left(-\frac{\gamma Rt}{\sqrt{1-\gamma^{2}R^{2}}}\right)(\kappa-t)^{2}
\end{split}
.\label{appeq:sc_saddle_4-5_q=00003D1}
\end{equation}
The free energy is (recall that $\gamma=1/\sqrt{1+\sigma^{2}}$)

\begin{equation}
G=\frac{1}{2(1-q)}\left(\Delta-\tilde{q}^{2}-2\tilde{R}R+\frac{1}{\Delta}\left\langle \left[\tilde{R}w^{0}+t\tilde{q}\right]_{+}^{2}\right\rangle _{t,w^{0}}\right)-\alpha\int_{-\infty}^{\kappa}DtH\bigg(-\frac{\gamma Rt}{\sqrt{1-\gamma^{2}R^{2}}}\bigg)(\kappa-t)^{2}.
\end{equation}

\subsubsection{Replica calculation of generalization with distribution-constraint}

\label{app:generalization_dist_const}

In this subsection, we will consider the case where student weights
are constrained to some \textit{prior} distribution $q_{s}(w_{s})$,
while the teacher obeys a distribution $p_{t}(w_{t}),$for an arbitrary
pair $q_{s},p_{t}$. We can write down the Gardner volume $V_{g}$
for generalization as in the capacity case (main text Eqn.2):

\begin{equation}
V_{g}=\frac{\int d\boldsymbol{w}_{s}\left[\prod_{\mu=1}^{P}\Theta\left(\text{sgn}\left(\frac{\boldsymbol{w}_{t}\cdot\boldsymbol{\xi}^{\mu}}{||\boldsymbol{w}_{t}||}+\eta^{\mu}\right)\frac{\boldsymbol{w}_{s}\cdot\boldsymbol{\xi}^{\mu}}{||\boldsymbol{w}_{s}||}-\kappa\right)\right]\delta(||\boldsymbol{w}_{s}||^{2}-N)\delta\bigg(\int dk\left(\hat{q}(k)-q(k)\right)\bigg)}{\int d\boldsymbol{w}_{s}\delta(||\boldsymbol{w}_{s}||^{2}-N)}.\label{appeq:gen_dc_vol}
\end{equation}

We treat the distribution constraint $q_{s}(w)$ similar to Section
\ref{app:capacity_dist_const}. The entropic part of the free energy
becomes

\begin{equation}
\begin{split}G_{0}= & -\frac{1}{2}\hat{q}_{1}+\frac{1}{2}\hat{q}q-\hat{R}R+\int_{-\infty}^{\infty}dwq_{s}(w)\lambda(w)+\left\langle \ln Z\right\rangle _{t,w_{t}}\\
Z= & \int\frac{dw}{\sqrt{2\pi}}\exp\left[-f(w)\right]\\
f(w)= & \frac{1}{2}(\hat{q}-\hat{q}_{1})w^{2}-(\hat{R}w_{t}+t\sqrt{\hat{q}})w+\lambda(w)
\end{split}
.
\end{equation}
At the limit $q\rightarrow1$, we make the following ansatz

\begin{equation}
\hat{R}=\frac{\tilde{R}}{1-q};\quad\hat{q}=\frac{u^{2}}{(1-q)^{2}};\quad\hat{q}-\hat{q}_{1}=\frac{\Delta}{1-q};\quad\lambda(w)=\frac{r(w)}{1-q}.
\end{equation}
Then

\begin{equation}
\begin{split}G_{0}= & \frac{1}{\left(1-q\right)}\left(-\frac{1}{2}u^{2}+\frac{1}{2}\Delta-\tilde{R}R+\int dwq_{s}(w)r(w)\right)+\langle\ln Z\rangle_{t,w_{t}}\\
f(w)= & \frac{1}{1-q}\left(\frac{1}{2}\Delta w^{2}-(\tilde{R}w_{t}+ut)w+r(w)\right)
\end{split}
\end{equation}
We can absorb $\frac{1}{2}\Delta w^{2}$ into the definition of $r(w)$,
$\frac{1}{2}\Delta w^{2}+r(w)\to r(w)$, and $0=\partial G_{0}/\partial\Delta$
gives the second moment constraint, $1=\int dwq_{s}(w)w^{2}$. 

Then,

\begin{equation}
\begin{split}G_{0}= & \frac{1}{\left(1-q\right)}\left(-\frac{1}{2}u^{2}-\tilde{R}R+\int dwq(w)r(w)\right)+\langle\ln Z\rangle_{t,w_{t}}\\
f(w)= & \frac{1}{1-q}\left(r(w)-(\tilde{R}w_{t}+ut)w\right)
\end{split}
.
\end{equation}
Next, we perform a saddle-point approximation on the log-term in $G_{0}$,
\begin{equation}
Z=\int\frac{dw}{\sqrt{2\pi}}\exp\left[-f(w)\right]\approx\exp\left[-f(w_{s})\right],
\end{equation}
where $w_{s}$ is the saddle-point value for the weight, and is determined
implicitly by 
\begin{equation}
r'(w_{s})=\tilde{R}w_{t}+ut.
\end{equation}
Note that $r'(w_{s})$ is now an induced random variable from random
variables $w_{t}$ and $t$. For later convenience, we rescale $r'(w_{s})$
to define a new random variable $z$,
\begin{equation}
z\equiv u^{-1}r'(w_{s})=t+u^{-1}\tilde{R}w_{t}\equiv t+\varepsilon w_{t},
\end{equation}
where we have also defined 
\begin{equation}
\varepsilon\equiv u^{-1}\tilde{R}.
\end{equation}
The induced distribution on $z$ is then 
\begin{equation}
\tilde{p}(z)=\int Dt\int dw_{t}p(w_{t})\delta(z-t-\varepsilon w_{t}).
\end{equation}
Now the entropic part becomes
\begin{equation}
G_{0}=\frac{1}{\left(1-q\right)}\left(-\frac{1}{2}u^{2}-\tilde{R}R+\int dwq_{s}(w)r(w)+\langle(\tilde{R}w_{t}+ut)w_{s}\rangle_{t,w_{t}}-\langle r(w_{s})\rangle_{t,w_{t}}\right).
\end{equation}
Integrate by parts,
\begin{equation}
\int dwq(w)r(w)=-\int dwQ(w)r'(w),
\end{equation}

\begin{equation}
\begin{split}\langle r(w_{s})\rangle_{t,w_{t}}= & \int Dtdw_{t}p_{t}(w_{t})r(w_{s})\\
= & \int dz\delta(z-t-\varepsilon w_{t})\int Dtdw_{t}p_{t}(w_{t})r(w_{s})\\
= & \int dz\tilde{p}(z)r(w_{s})\\
= & -\int dz\tilde{P}(z)r'(w_{s})
\end{split}
.\label{appeq:113}
\end{equation}
Now $0=\partial G/\partial r'(w_{s})$ gives

\begin{equation}
Q(w_{s})=\tilde{P}(z).
\end{equation}
 which implicitly determines $w_{s}(z).$ 

Next,
\begin{equation}
0=\frac{\partial G}{\partial u}\Rightarrow u=\langle w_{s}(z)t\rangle_{t,w_{t}},
\end{equation}
\begin{equation}
0=\frac{\partial G}{\partial\tilde{R}}\Rightarrow R=\langle w_{s}(z)w_{t}\rangle_{t,w_{t}}.
\end{equation}

The free energy then simplifies to 
\begin{equation}
G=\frac{u^{2}}{2\left(1-q\right)}+\alpha G_{1}.
\end{equation}
The energetic part as $q\to1$ becomes (same as the unconstrained
and sign-constrained case)

\begin{equation}
G_{1}=-\frac{1}{1-q}\int_{-\infty}^{\kappa}DtH\bigg(-\frac{\gamma Rt}{\sqrt{1-\gamma^{2}R^{2}}}\bigg)(\kappa-t)^{2}.
\end{equation}
The remaining two saddle point equations are (1) the vanishing log-Gardner
volume and (2) $0=\partial G/\partial R$:
\begin{equation}
\frac{1}{2}u^{2}=\alpha\int_{-\infty}^{\kappa}DtH\bigg(-\frac{\gamma Rt}{\sqrt{1-\gamma^{2}R^{2}}}\bigg)(\kappa-t)^{2},
\end{equation}
\begin{equation}
\varepsilon u=\alpha\gamma\sqrt{\frac{2}{\pi}}\sqrt{1-\gamma^{2}R^{2}}\int_{-\tilde{\kappa}}^{\infty}Dt\bigg(\tilde{\kappa}+t\bigg).
\end{equation}

In summary, the order parameters $\left\{ R,\kappa,u,\varepsilon\right\} $
can be determined from a set of self-consistency equations:

\begin{equation}
\begin{split}u= & \langle w_{s}(z)t\rangle_{t,w_{t}}\\
R= & \langle w_{s}(z)w_{t}\rangle_{t,w_{t}}\\
\frac{1}{2}u^{2}= & \alpha\int_{-\infty}^{\kappa}DtH\bigg(-\frac{\gamma Rt}{\sqrt{1-\gamma^{2}R^{2}}}\bigg)(\kappa-t)^{2}\\
\varepsilon u= & \frac{2\alpha\gamma}{\sqrt{2\pi}}\sqrt{1-\gamma^{2}R^{2}}\int_{-\tilde{\kappa}}^{\infty}Dt\bigg(\tilde{\kappa}+t\bigg)
\end{split}
,
\end{equation}

where we have introduced $\tilde{\kappa}=\kappa/\sqrt{1-\gamma^{2}R^{2}}$,
an auxiliary normal variable $t\sim\mathcal{N}(0,1)$, and an induced
random variable $z\equiv t+\varepsilon w_{t}$ with induced distribution

\begin{equation}
\tilde{p}(z)=\int Dt\int dw_{t}p_{t}(w_{t})\delta(z-t-\varepsilon w_{t}).\label{appeq:induced_dist}
\end{equation}
Note that $w_{s}(z)$ can be determined implicitly by equating the
CDF of the induced variable $z$ and the distribution that the student
is constrained to: 
\begin{equation}
Q(w_{s})=\tilde{P}(z).
\end{equation}

\subsubsection*{Examples}

(1) Lognormal distribution

In the following, we solve $w_{s}(z)$ explicitly from the CDF equation
$Q(w_{s})=\tilde{P}(z)$. For a lognormal teacher,

\begin{equation}
p_{t}(w_{t})=\frac{1}{w_{t}}\frac{1}{\sqrt{2\pi}\sigma}\exp\left\{ -\frac{(\ln w_{t}-\mu)^{2}}{2\sigma^{2}}\right\} .
\end{equation}
The second moment constraint implies $\mu=-\sigma^{2}.$

The induced CDF of $z$ is
\begin{equation}
\tilde{P}(z)=\int_{-\infty}^{z}dz'\int_{-\infty}^{\infty}Dt\int_{0}^{\infty}dw_{t}p_{t}(w_{t})\delta(z'-t-\varepsilon w_{t}).
\end{equation}
Let $x=(\ln w-\mu)/\sigma,$

\begin{equation}
\begin{split}\tilde{P}(z)= & \int_{-\infty}^{z}dz'\int_{-\infty}^{\infty}Dt\int_{-\infty}^{\infty}Dx\delta(z'-t-\varepsilon e^{\mu+\sigma x})\\
= & \int_{-\infty}^{\infty}DxH(\varepsilon e^{\mu+\sigma x}-z)
\end{split}
.
\end{equation}
Now the CDF of $w_{s}$ is

\begin{equation}
\begin{split}Q_{s}(w_{s})= & \int_{-\infty}^{w_{s}}q_{s}(w)dw=H\left(-\frac{\ln w_{s}-\mu}{\sigma}\right)\end{split}
.
\end{equation}
Therefore, equating $\tilde{P}(z)$ and $Q_{s}(w_{s})$:
\begin{equation}
\int_{-\infty}^{\infty}DxH(\varepsilon e^{\mu+\sigma x}-z)=H\left(-\frac{\ln w_{s}-\mu}{\sigma}\right),
\end{equation}

We can solve for \textbf{$w_{s}(z)$} by (recall\textbf{ $z\equiv t+\varepsilon w_{t}$})
\begin{equation}
w_{s}(z)=\exp\left\{ \mu+\sigma H^{-1}\left(\int DxH(z-\varepsilon e^{\mu+\sigma x})\right)\right\} .
\end{equation}

Or in terms of error functions
\begin{equation}
w_{s}(z)=\exp\left\{ \mu+\text{\ensuremath{\sqrt{2}\sigma}erf}^{-1}\left(\int Dx\text{erf}\left(\frac{\varepsilon e^{\mu+\sigma x}-z}{\sqrt{2}}\right)\right)\right\} .
\end{equation}

We can also calculate the initial overlap (before any learning):
\begin{equation}
R_{0}=\left\langle \boldsymbol{w}_{t}\cdot\boldsymbol{w}_{s}\right\rangle _{p_{t}q_{s}}=e^{2\mu+\sigma^{2}}=e^{-\sigma^{2}}.
\end{equation}

(2) Uniform distribution 

Assuming that both the teacher and the student have a uniform distribution
in range $[0,\sigma].$

The second moment constraint fixes $\sigma=\sqrt{3}.$

We can solve (as in the lognormal example above), 

\begin{equation}
w_{s}(z)=\frac{1}{\varepsilon}\int_{-\infty}^{z}dz'\left(H(z'-\varepsilon\sigma)-H(z')\right).
\end{equation}

(3) Half-normal distribution

Assuming that both the teacher and the student has a half-normal distribution
$\frac{2}{\sqrt{2\pi}\sigma}\exp\left\{ -\frac{w^{2}}{2\sigma^{2}}\right\} $.

The second moment constraint fixes $\sigma=1$, and 

\begin{equation}
w_{s}(z)=\sigma H^{-1}\left\{ \frac{1}{2}-\int_{-\infty}^{\frac{z}{\sqrt{1+\sigma^{2}\varepsilon^{2}}}}DtH(-\sigma\varepsilon t)\right\} .
\end{equation}

\subsubsection*{Arbitrary number of synaptic subpopulations}

Just like in the case of Section \ref{app:capacity_M_pop}, we can
generalize our theory above to incorporate distribution constraints
with an arbitrary number of synaptic subpopulations. Let's consider
a student perceptron with $M$ synaptic populations indexed by $m$,
$\boldsymbol{w}^{m}$, such that each $w_{i}^{m}$ satisfies its own
distributions constraints $w_{i}^{m}\sim Q_{m}(w^{m})$. We denote
the overall weight vector as $\boldsymbol{w}\equiv\{\boldsymbol{w}^{m}\}_{m=1}^{M}\in\mathbb{R}^{N\times1}$.
The total number of weights is $N=\sum_{m=1}^{M}N_{m}$, and we denote
the fractions as $g_{m}=N_{m}/N$. Since the derivation is similar
to that of Section \ref{app:capacity_M_pop} and Section \ref{app:generalization_dist_const},
we will only present the results here.

As before, the order parameters $\left\{ R,\kappa,u,\varepsilon\right\} $
can be determined from a set of self-consistency equations:

\begin{equation}
\begin{split}u= & \sum_{m}g_{m}\langle w^{m}(z)t\rangle_{t,w_{t}}\\
R= & \sum_{m}g_{m}\langle w^{m}(z)w_{t}\rangle_{t,w_{t}}\\
\frac{1}{2}u^{2}= & \alpha\int_{-\infty}^{\kappa}DtH\bigg(-\frac{\gamma Rt}{\sqrt{1-\gamma^{2}R^{2}}}\bigg)(\kappa-t)^{2}\\
\varepsilon u= & \frac{2\alpha\gamma}{\sqrt{2\pi}}\sqrt{1-\gamma^{2}R^{2}}\int_{-\tilde{\kappa}}^{\infty}Dt\bigg(\tilde{\kappa}+t\bigg)
\end{split}
,
\end{equation}

where $\tilde{\kappa}=\kappa/\sqrt{1-\gamma^{2}R^{2}}$, $t\sim\mathcal{N}(0,1)$.
and an induced random variable $z\equiv t+\varepsilon w_{t}$ with
induced distribution the same as Eqn.\ref{appeq:induced_dist}.

Note that every $w^{m}(z)$ can be determined by equating the CDF
of the induced variable $z$ and the $m$-th distribution that $w^{m}(z)$
is constrained to: 
\begin{equation}
Q_{m}(w^{m})=\tilde{P}(z).
\end{equation}

\subsubsection{Sparsification of weights in sign-constraint learning}

\label{app:generalization_sparsification}

\begin{figure}
\centering{}\includegraphics[scale=0.16]{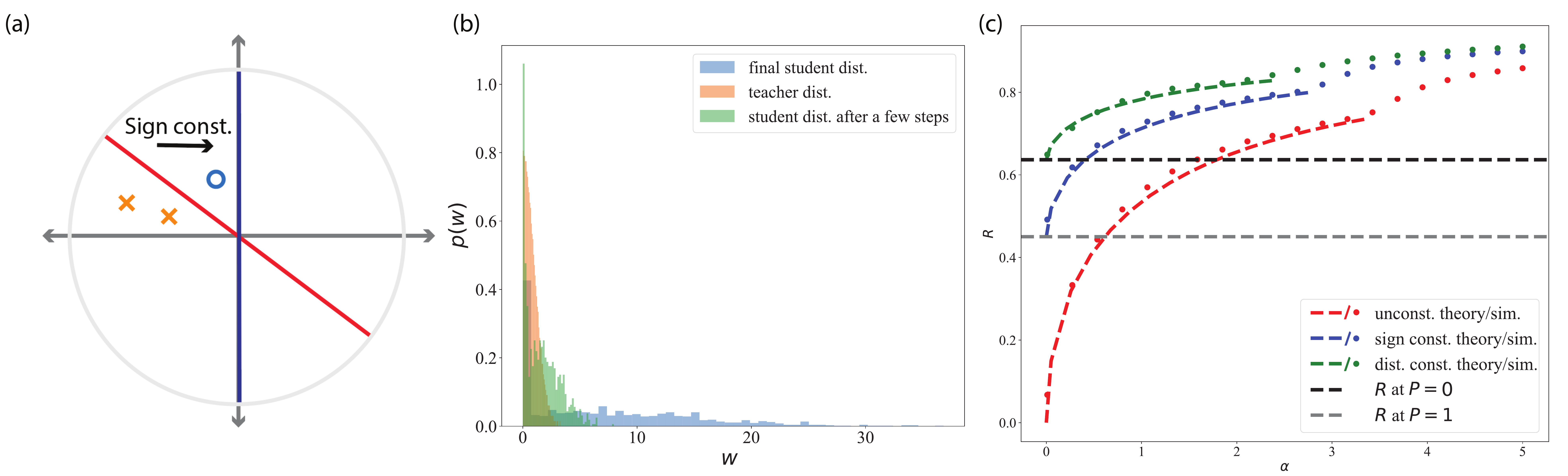}\caption{\label{appfig:sparsification} Sparsification of weights in sign-constraint
learning. (a) An illustration of weight sparsification. In this schematic,
the perceptron lives on this $1$-dimensional circle and $N=2$. Red
line denotes the hyperplane orthogonal to the perceptron weight before
sign-constraint, crosses and circles indicate examples in different
classes. Sign-constraint pushes the weights to the first quadrant,
which zeros half of the weights on average. Blue line indicates the
hyperplane obtained after the sign-constraint. (b) Sparsification
of weights due to max-margin training. After only a few iterations,
nearly half of the student weights are set to zero, and the distribution
deviates significantly from the teacher's distribution. (c) Teacher-student
overlap as a function of load $\alpha$ for different learning paradigms.
Dashed lines are from theory, and dots are from simulation. Note the
horizontal dashed lines show the initial drop in overlap from zero
example and to just a single example. In this case teacher has nonzero
noise, $\gamma=0.85$. }
\end{figure}

For unconstrained weights, max-margin solutions are considered beneficial
for generalization particularly for small size training sets. As a
first step toward biological plausibility, one can try to constraint
the sign of individual weights during learning (e.g., excitatory or
inhibitory). In the generalization error setup, we can impose a constraint
that the teacher and student have the same set of weight signs. Surprisingly,
we find both analytically and numerically that if the teacher weights
are not too sparse, the max-margin solution generalizes poorly: after
a single step of learning (with random input vectors), the overlap,
$R$, drops substantially from its initial value $R_{0}$ (by a factor
of $\sqrt{2}$ for a half-Gaussian teacher, see the blue curves in
Fig.\ref{appfig:sparsification}(a). 

We can verify this by calculating $R_{0}$ in two different ways.
As an example, in the following we consider the case where both the
teacher and student have half-normal distributions. 

(1) By definition, the overlap is $R=\frac{\boldsymbol{w}_{s}\cdot\boldsymbol{w}_{t}}{\left\Vert \boldsymbol{w}_{s}\right\Vert \left\Vert \boldsymbol{w}_{t}\right\Vert }$.
Since $\boldsymbol{w}_{s}$ and $\boldsymbol{w}_{t}$ are uncorrelated
before learning ($\alpha=0$), the initial overlap is then $R_{0}=\frac{\left\langle w_{s}\right\rangle \left\langle w_{t}\right\rangle }{||\boldsymbol{w}_{s}||\boldsymbol{w}_{t}||}$$=\frac{2}{\pi};$

(2) Take the $\alpha\to0$ limit in Eqn.\ref{appeq:sc_saddle_1-3_qneq1}
and Eqn.\ref{appeq:sc_saddle_4-5_qneq1} and calculate $R_{0+}=\lim_{\alpha\to0+}R(\alpha)$
$=\frac{\sqrt{2}}{\pi}$. 

Therefore, in this example $R_{0+}=R_{0}/\sqrt{2}.$

The source of the problem is that due to the sign constraint, max-margin
training with few examples yields a significant mismatch between the
student and teacher weight distributions. After only a few steps of
learning, half of the student's weights are set to zero, and the student's
distribution, $p(w_{s})=\frac{1}{2}\delta(0)+\frac{1}{\sqrt{2\pi}}\exp\{-\frac{w_{s}^{2}}{4}\}$,
deviates significantly from the teacher's half-normal distribution
(Fig.\ref{appfig:sparsification}(b)).

\subsubsection{Noisy teacher}

\label{app:generalization_noise}

We generate examples $\{\boldsymbol{\xi}^{\mu},\zeta^{\mu}\}_{\mu=1}^{P}$
from a teacher perceptron, $\boldsymbol{w}_{t}\in\mathbb{R}^{N}$:
$\zeta^{\mu}=\text{sgn}(\boldsymbol{w}_{t}\cdot\boldsymbol{\xi}^{\mu}/||\boldsymbol{w}_{t}||+\eta^{\mu})$,
where $\eta^{\mu}$ is input noise and $\eta^{\mu}\sim\mathcal{N}(0,\sigma^{2})$.
In this subsection we present additional numerical results for the
case when $\sigma\neq0$. As in previous sections, we define the noise
level parameter $\gamma=1/\sqrt{1+\sigma^{2}}.$

\begin{figure}
\centering{}\includegraphics[scale=0.17]{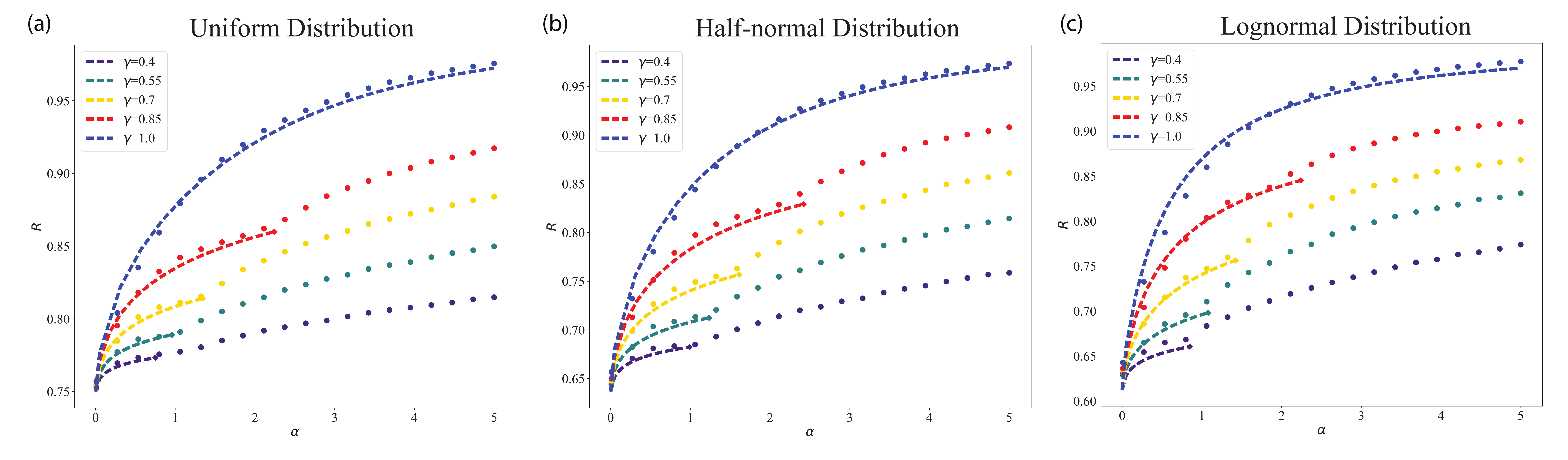}\caption{\label{appfig:vary_noise}Generalization (measured by overlap) performance
for different distributions and different noise levels in fixed prior
learning. From left to right: uniform, half-normal, and lognormal
distribution. In all cases the student is constrained to have the
same distribution as that of the teacher's. Dashed lines are from
theory and dots are from DisCo-SGD simulation.}
\end{figure}

Our theory's prediction is confirmed by numerical simulation for a
wide range of teacher noise level $\gamma$ and teacher weight distributions
$P_{t}(w_{t})$. We find that distribution-constrained learning performs
consistently better all the way up to capacity (capacity in this framework
is due to teacher noise). For illustration, in Fig.\ref{appfig:vary_noise}
we show theory and simulation for fixed prior learning of three different
teacher distributions: uniform, half-normal, and lognormal.

\subsection{DisCo-SGD simulations}

\label{app:DiscoSGD}

\subsubsection*{Avoid vanishing gradients}

Note that we often observe a vanishing gradient in DisCo-SGD when
we choose a constant learning rate $\eta_{1}$, and in such cases
the algorithm tends to find poor margin $\kappa$ which deviates from
the max-margin value predicted from the theory. We find that scaling
$\eta_{1}$ with the standard deviation of the gradient solves this
problem:

\begin{equation}
\eta_{1}=\eta_{1}^{0}/\text{std}\left(\sum_{\mu}\xi_{i}^{\mu}(\hat{\zeta}^{\mu}-\zeta^{\mu})\right),
\end{equation}

where the standard deviation is computed across the synaptic index
$i$ and $\eta_{1}^{0}$ is a constant. 

\subsubsection*{Mini-batches}

For the capacity simulations, we always use full-batch in the SGD
update, so it is in fact simply gradient descent. However, in the
case of generalization, we find that training with mini-batches improves
the generalization performance, since it acts as an source of stochasticity
during training. In main text Fig.5 we use mini-batch size $B=0.8P$
($80\%$ of examples are used for each SGD update).

When we vary teacher's noise level, we find that scaling $B$ with
$\gamma$ improves the quality of the solutions, as measured by the
generalization performance (or equivalently, the teacher-student overlap).
Generally, the more noisy the teacher is, the smaller the mini-batches
should be. This is because smaller mini-batch size corresponds to
higher stochasticity, which helps overcoming higher teacher noise. 

\subsubsection*{Parameters}

All the capacity simulations are performed with the following parameters
$N=1000,\eta_{1}^{0}=0.01,\eta_{2}=0.6,t_{max}=10000,$ where $t_{max}$
is the maximum number of iterations of the DisCo-SGD algorithm. 

All results are averaged over 300 realizations.

In main text Fig.4, the experimental \cite{levy2012spatial} parameters
are $g_{E}=45.8\%,\sigma_{E}=0.833,\sigma_{I}=0.899$.

In main text Fig.5(a): We show the teacher-student overlap as a function
of $\alpha$. Dots are simulations performed with series of student
distribution from $\sigma_{s}=0.1$ to $\sigma_{s}=1.4$, where the
teacher distribution sits in the middle of this range, $\sigma_{t}=0.7$.
Each such simulation is performed with fixed $\sigma_{s}$ and varying
load $\alpha\in[0.05,2.5].$ In main text Fig.5(b): we show the empirical
weight distributions found by unconstrained perceptron learning for
$\alpha\in[0.05,10]$. In main text Fig.5(c) we show optimal student
distribution for $\alpha\in[0.05,2.5]$. Note that optimal prior learning
approaches the teacher distribution much faster than unconstrained
learning. 

All the generalization DisCo-SGD simulations are performed with the
same parameter as in the capacity DisCo-SGD simulations, but with
two additional parameter: teacher's noise level $\gamma$ and SGD
mini-batch size $B$. 

For the simulations in Fig.\ref{appfig:vary_noise} we use

$\gamma=0.4,B=0.2P;\gamma=0.55,B=0.4P;\gamma=0.7,B=0.6P;\gamma=0.85,B=0.8P;\gamma=1.0,B=P$
(noiseless case).

\subsection{Replica symmetry breaking}

\label{app:Replica-symmetry-breaking}

\subsubsection{Bimodal distributions}

In deriving the capacity formula, we have assumed replica-symmetry
(RS). It is well-known that replica-symmetry breaking occurs in the
Ising perceptron \cite{penney1993weight,bouten1998learning}, so it
is natural to ask to what extent our theory holds when approaching
the Ising limit. Let's consider a bimodal distribution with a mixture
of two normal distributions with non-zero mean centered around zero,

\[
p(w)=\frac{1}{2}\mathcal{N}(-\mu,\sigma)+\frac{1}{2}\mathcal{N}(\mu,\sigma)
\]
The second moment constraint requires $\mu^{2}+\sigma^{2}=1$.

We can gradually decrease the Gaussian width $\sigma$, or equivalently
$\mu=\sqrt{1-\sigma^{2}}$ (which we call `separation' in the following)
and compare the capacity theoretically predicted by the RS theory
and numerically found by the DisCo-SGD algorithm.

\begin{figure}
\centering{}\includegraphics[scale=0.27]{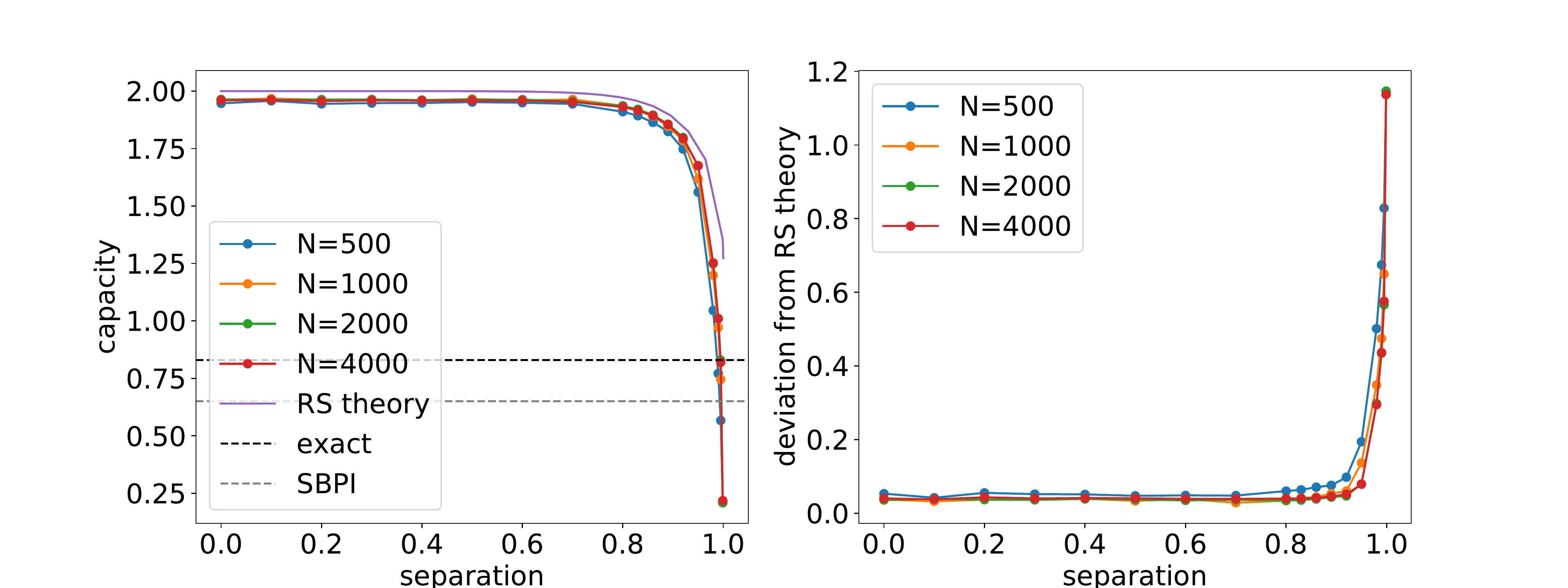}\caption{\label{appfig:replica_symmetry_breaking}Left: Capacity as a function
of separation for different size perceptrons. Dots are from DisCo-SGD
simulations and the `RS theory' line is from our theory. Exact values
for Ising perceptron and state-of-the-art numerical values are included
as well. Right: Deviation from the RS theory as a function of separation.
This is the same as subtracting the simulation values from the theoretical
predictions in the left figure. }
\end{figure}

\begin{figure}
\centering{}\includegraphics[scale=0.27]{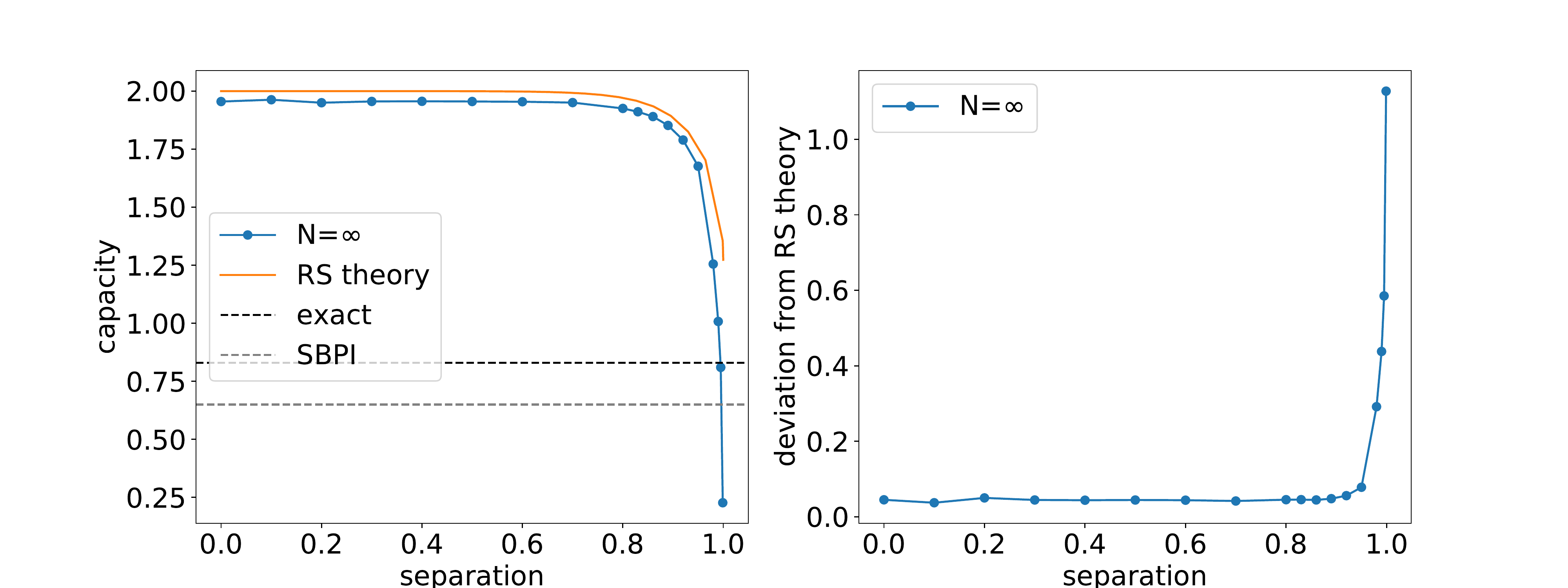}\caption{\label{appfig:finite_size_effects}Finite size effects. Left/Right:
we extrapolate simulation values in Fig.\ref{appfig:replica_symmetry_breaking}
Left/Right to infinite $N$. }
\end{figure}

In Fig.\ref{appfig:replica_symmetry_breaking} we can see that the
simulation agrees well with the RS theory until one gets very close
to the Ising limit ($\mu=1$). To understand finite size effects,
we extrapolate to the infinite size limit ($N\to\infty$) in Fig.\ref{appfig:finite_size_effects},
and found that the deviation from RS theory has a sharp transition
near $\mu=1$, marking the breakdown of the RS theory.

\subsubsection*{Ising perceptron}

It is also interesting to compare our distribution-constrained RS
theory to the unconstrained RS theory. In this Ising limit,

\begin{equation}
q(w)=\frac{1}{2}\delta(w-1)+\frac{1}{2}\delta(w+1),
\end{equation}

and CDF
\begin{equation}
Q(w)=\frac{1}{2}\Theta(w-1)+\frac{1}{2}\Theta(w+1).
\end{equation}

Equating $Q(w)$ with the normal CDF $P(x)$ and solve for $w(x)$,
we find $w(x)=\text{sgn}(x)$. Then $dw/dx=2\delta(x)$ and $\left\langle \frac{dw}{dx}\right\rangle _{x}=\frac{2}{\sqrt{2\pi}}$.
Therefore,

\begin{equation}
\lim_{Ising}\alpha_{c}(\kappa=0)=\frac{4}{\pi},
\end{equation}

which is exactly the same as the prediction from the unconstrained
RS theory \cite{penney1993weight,bouten1998learning}. In contrast,
the exact capacity of Ising perceptron with replica-symmetry breaking
is $\alpha_{c}\approx0.83$. For comparison, we have included these
values in Fig.\ref{appfig:finite_size_effects}(a), as well as the
capacity found by the state-of-the-art supervised learning algorithm
(Stochastic Belief Propagation, SBPI \cite{baldassi2007efficient})
for Ising perceptron.

\subsubsection{Sparse distributions}

\begin{figure}
\centering{}\includegraphics[scale=0.23]{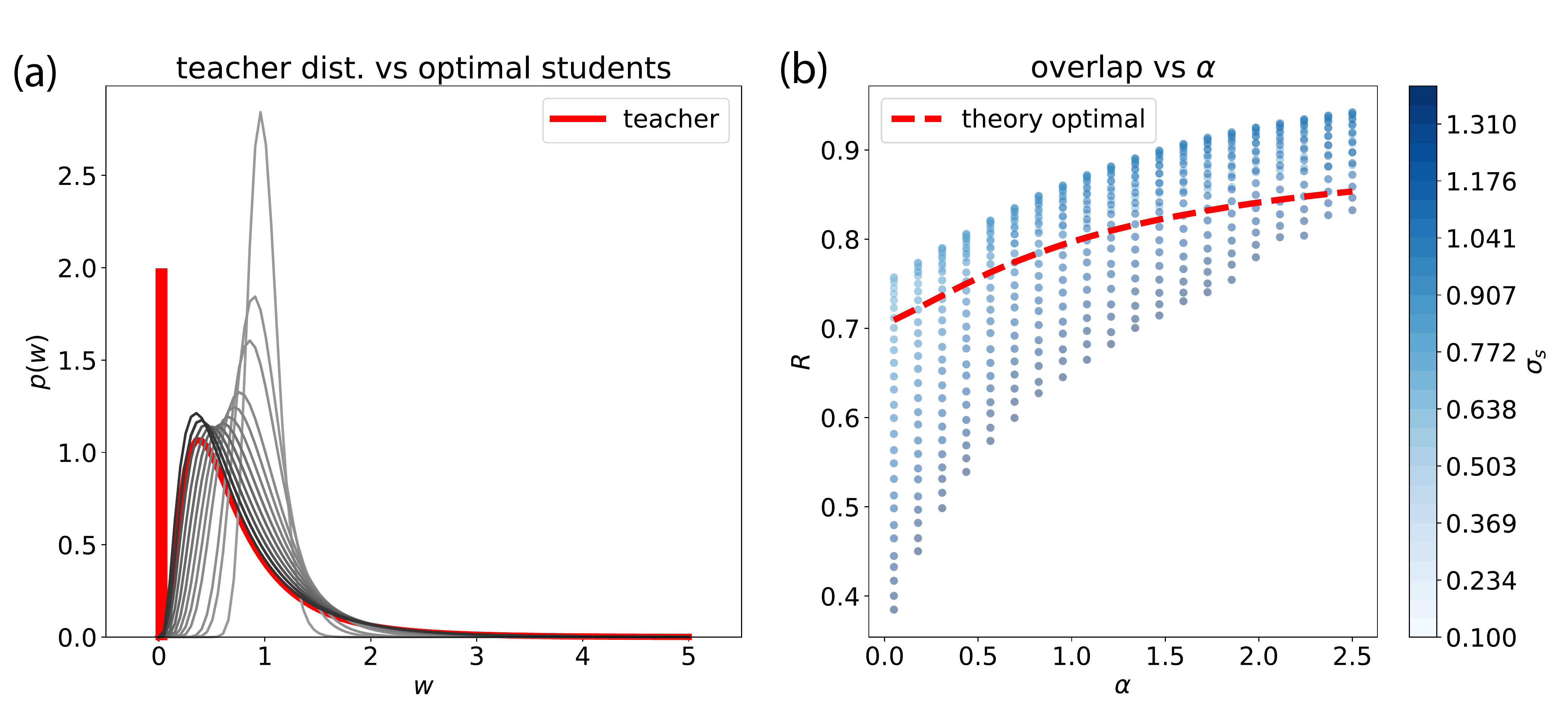}\caption{\label{appfig:sparse_teacher}Optimal student prior distribution as
a function of $\alpha$. (a) Gray curves correspond to a series of
optimal student distributions as a function of $\alpha$, with the
darker color representing larger $\alpha$. Red is teacher distribution.
(b) Overlap as a function of $\alpha$ for different student priors.
Red dashed line is the optimal overlap calculated from our replica-symmetric
theory. Dots are from DisCo-SGD simulations. For the same $\alpha$,
different color dots represent different overlaps obtained from simulations
with different $\sigma_{s}$. }
\end{figure}

For a teacher with sparse distribution, $p(w_{t})=(1-\rho)\delta(w_{t})+\text{\ensuremath{\frac{\rho}{\sqrt{2\pi}\sigma_{t}w_{t}}\exp\left\{ -\frac{(\ln w_{t}-\mu_{t})^{2}}{2\sigma_{t}^{2}}\right\} }. }$We
found that the simulations start to deviate from the theory, and the
reason might be due to replica symmetry breaking. In Fig.\ref{appfig:sparse_teacher},
we use the optimal prior learning paradigm similar to main text Fig.5.
We see that our RS theory no longer gives accurate prediction of overlap
in this case.

\end{document}